\UseRawInputEncoding

\documentclass[preprint,11pt]{elsarticle}

\usepackage[section]{placeins}
\usepackage{subfigure}
\usepackage[section]{placeins}
\usepackage{epsfig}

\usepackage{amssymb}

	\usepackage{siunitx}

	\usepackage{xcolor}

	\usepackage{color,soul}

	\usepackage{ulem}

	\usepackage{multirow, tabularx, graphicx, array, footnote, multicol}
	\usepackage{booktabs}
	\usepackage{threeparttable}

	\usepackage{amsmath}

	\usepackage{scalerel}
	
	\journal{Journal name}

\begin{document}

	\begin{frontmatter}



		\title{ When carbon impurities trigger the synthesis of alpha boron at high pressure and high temperature} 

	\author[label1,label2,label4]{Amrita Chakraborti}
	\ead{Amrita.Chakraborti@uni-bayreuth.de}

	\author[label1]{Yeonsoo Cho}
	\author[label1]{Jelena Sjakste}
	\author[label2]{Benoit Baptiste}
	\author[label3]{Laura Henry}
	\author[label3]{Nicolas Guignot}
	\author[label2]{Yann Le Godec}
	\ead{yann.le_godec@sorbonne-universite.fr}
	\author[label1]{Nathalie Vast}
	\ead{nathalie.vast@polytechnique.edu}

	\address[label1]{Laboratoire des Solides Irradiés, École Polytechnique, CEA/DRF/IRAMIS, CNRS UMR 7642, Institut Polytechnique de Paris, 28 Route de Saclay, F-91128, Palaiseau Cedex, France}
	\address[label2]{Institut de Minéralogie, de Physique des Matériaux et de Cosmochimie (IMPMC), Sorbonne Université, UMR CNRS 7590, Muséum National d'Histoire Naturelle, IRD UMR 206, 4 Place Jussieu, 75005, Paris, France}
	\address[label3]{Synchrotron SOLEIL, Gif-sur-Yvette, France}
	\address[label4]{Currently at Bayerisches Geoinstitut, Universität Bayreuth, Bayreuth D95440, Germany}

	\begin{abstract}
	The role of carbon in the formation of $\alpha$~boron at high pressure and high temperature (HPHT) has been investigated by combining HPHT experiments and density functional theory (DFT) calculations. Starting from $\beta$~rhombohedral or amorphous boron and amorphous carbon at 5~GPa, the $\alpha$~boron phase has been repeatedly observed between 1473~K and 2273~K, at temperatures that are much higher than those reported in the phase diagram of boron. The DFT investigation of the effect of carbon insertion into the $\alpha$~rhombohedral boron atomic structure on the formation enthalpy, volume and optical properties shows that only the (B$_{11}$C) substituted icosahedron defect accounts for the slight volume contraction observed in $\alpha$~boron. It is also compatible with the observed red colour of $\alpha$~boron crystals, and consistently has a very low formation energy. Our calculations also provide a calibration tool to evaluate the carbon concentration in the samples. Finally, the synthesis temperature is found to be an easy means to tune the carbon concentration on demand.

	\end{abstract}


			\begin{highlights}
			\item Formation of $\alpha$~boron from other boron phases treated with amorphous carbon at HPHT conditions
			\item Theoretical modeling of carbon insertion into the atomic structure of $\alpha$~boron accounting for the observed physical properties
                        \item Carbon impurities found to trigger the synthesis of $\alpha$~boron
			\end{highlights}

			\begin{keyword}


	alpha boron \sep High Pressure High Temperature synthesis \sep boron phase diagram \sep carbon impurities 

	\end{keyword}

	\end{frontmatter}


	\section{Introduction}
	\label{sec:Intro}

\begin{figure}[ht!]
	\centering
		\subfigure[Boron carbide]{\includegraphics[height=5cm]{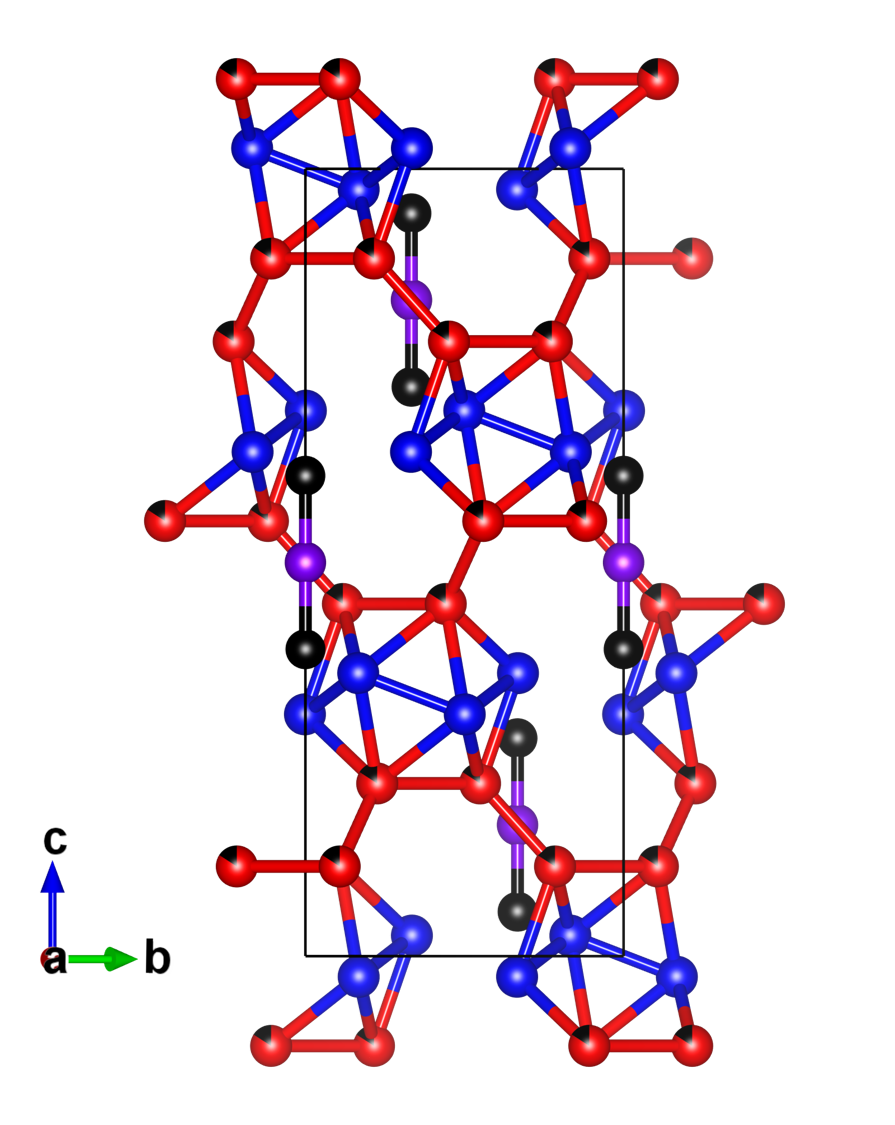} \label{fig:B4C}}
		\subfigure[$\alpha$~boron]{\includegraphics[height=5cm]{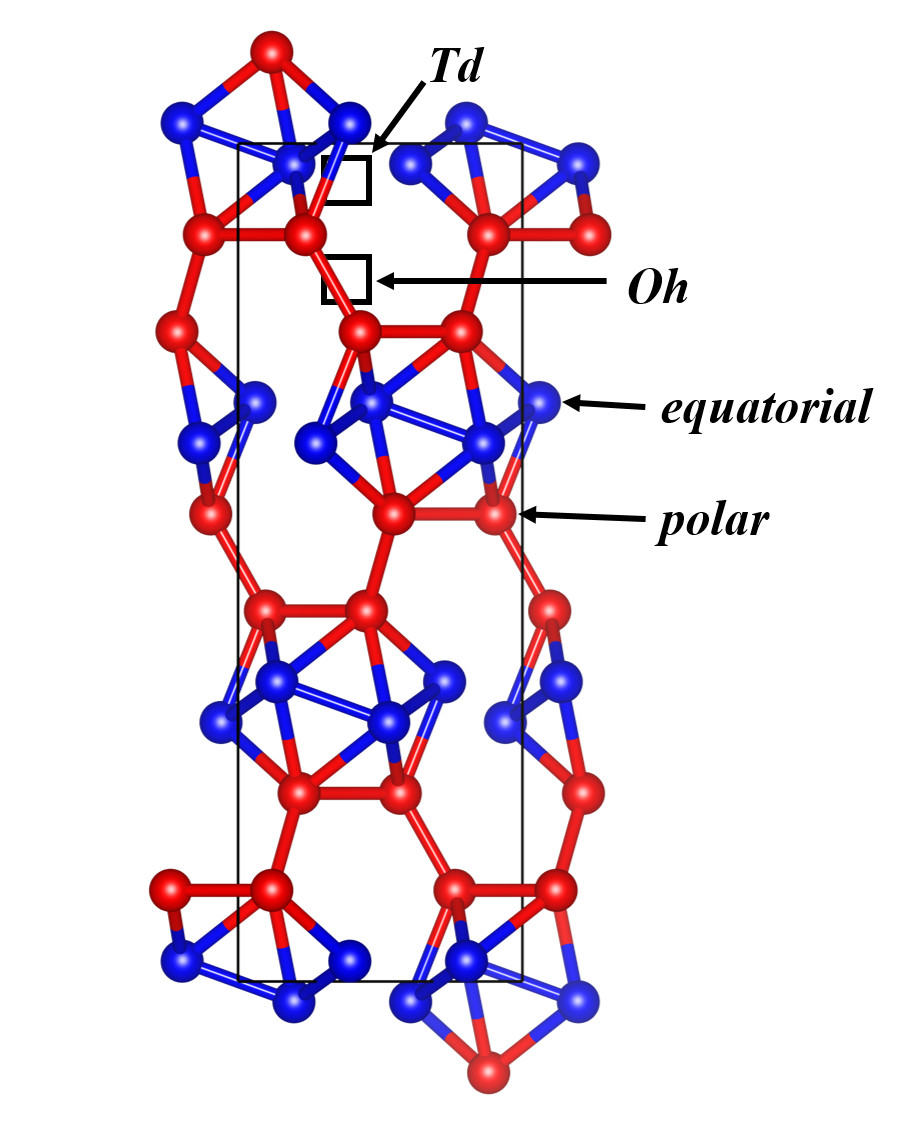} \label{fig:alpha_boron}}
		\subfigure[$\beta$ boron]{\includegraphics[height=5cm]{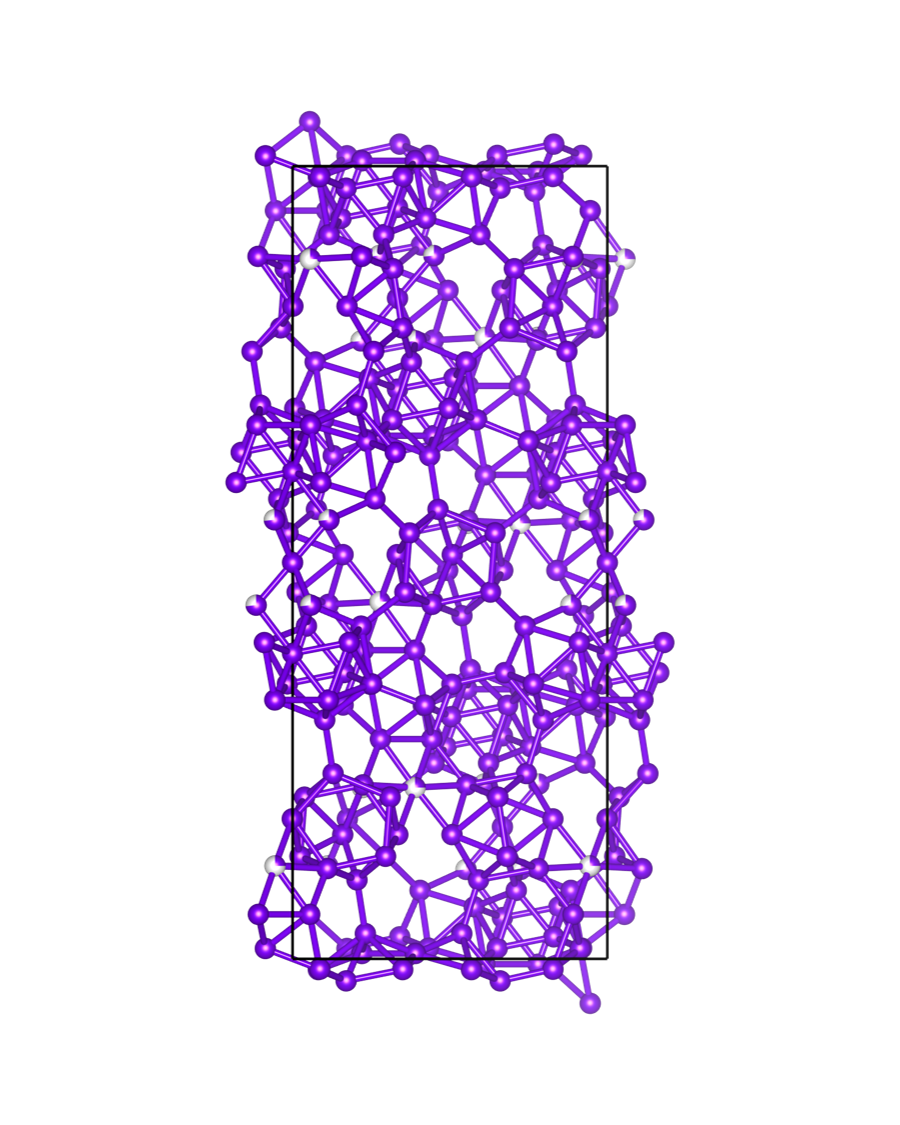} \label{fig:beta_boron}}
	\caption{\label{fig:boron_comparison} 
		\textit{R$\overline 3 $m} hexagonal conventional unit cells (u.c.) of (a) boron carbide, (b) $\alpha$~boron and (c) $\beta$ boron~\cite{Chakraborti:Note:2022:Figure_software,Momma:2011}. $\alpha$~boron and boron carbide crystal structures only differ by the presence of C-B-C chains in the intericosahedral space, and by the substitution of one B atom by one C atom in the polar site of each icosahedron.
Red balls represent boron atoms in the polar site of the icosahedra of $\alpha$~boron, which is the main substitution site for carbon discussed in the main text. Blue balls are equatorial boron atoms in $\alpha$~boron and in boron carbide. Violet balls represent chain-center boron atoms in boron carbide as well as all of the boron atoms of $\beta$~boron. Carbon atoms are represented as black balls, and the substitutional disorder of the carbon atom in the polar site of the icosahedra in boron carbide is represented by black-and-red balls.  
$\beta$~boron hexagonal u.c. has been assumed to contain 315 atoms~\cite{Amberger:1981,Hughes:1963,Geist:1970,Cho:Note:2022:beta_structure,Slack:1988,Hiroto:2020}, while $\alpha$~boron and boron carbide u.c. contain respectively 36 and 45 atoms.
	}
\end{figure}

Alpha rhombohedral boron ($\alpha$~boron) 
	has potential uses in a variety of fields, notably nuclear and space applications. The Vickers hardness of $\alpha$~boron is as high as 42~GPa~\cite{Amberger:1981}, making it a superhard material while the density is estimated to be 2.460~g/cm$^{3}$~\cite{Amberger:1981}. The damage under irradiation~\cite{Srour:2013} of crystals containing (B$_{12}$) or (B$_{11}$C) icosahedra (figure~\ref{fig:boron_comparison}), such as $\beta$~boron~\cite{Favia:1996}, boron carbide~\cite{Stoto:1985,Solozhenko:2004,Domnich:2011,Jay:2019}, or (B$_{12}$)P$_2$ and (B$_{12}$)As$_2$~\cite{Carrard:1995}, is much less severe than in other materials with a cubic crystal structure -~such as silicon~- or with a hexagonal crystal structure, such as gallium nitride or boron nitride. 
	The atoms of the icosahedra, displaced by irradiation, return to their initial position by a self-healing mechanism~\cite{Emin:2004,Emin:2006,Slack:2014}. 

	Interestingly, $\alpha$~boron is expected to combine this irradiation resistance with semi-conducting electronic properties and a direct band gap of $\sim$~2~eV~\cite{Horn:1959,Golikova:1979,Werheit:1991c,Shirai:2009}, making it suitable for space applications~\cite{Horn:1959,Shirai:2009}.
	This is in wide contrast with $\beta$~boron and boron carbide, in which the disorder of the atomic structure leads to a large number of electronic defect levels in the bandgap~\cite{Favia:1996,Jay:2019,Neft:1965,Werheit:1991b,Werheit:2009}.

	The useful electronic properties coupled to the natural abundance of the $^{10}$B isotope that reacts with neutrons, could result in applications such as solid state neutron counters~\cite{Cocks:2015}, and potentially facilitate the production of individual dosimeters in the nuclear industry. 
	However, so far, there is no established process for mass production of $\alpha$~boron, which necessitates the search for new synthesis processes. 

The methods generally used to synthesise $\alpha$~boron are incompatible with the hygiene and safety standards recommended for the manufacture on an industrial scale \cite{Albert:2009, McCarty:1958}. A preparation method 
of $\alpha$~boron crystals has been described in the literature from the toxic and corrosive reagent BI$_{3}$, on a surface heated between 1073~K and 1273~K~\cite{McCarty:1958}. The use of BBr$_{3}$ or BCl$_{3}$ reagents would also suffer from the same drawbacks linked to the dangers of using boron halides, namely, they are corrossive and toxic. Other known routes for the 
growth of $\alpha$~boron are expensive since they use noble metals and/or heavy metals such as platinum~(Pt) or palladium~(Pd), with very low yields leading to crystals having a size of only a few hundred micrometers~\cite{Tallant:1989,Polian:2008}.

	       One decade ago, syntheses of $\alpha$~boron at high pressures between 6~GPa and 11~GPa, starting from $\beta$~rhombohedral boron, have also been reported in literature~\cite{Parakhonskiy:2011b}. This $\alpha$~boron preparation must be done in presence of finely divided platinum (Pt), and in a Pt chamber, or even in gold (Au). Since the implementation of a reactor consisting of these noble metals is not possible on an industrial scale, this synthesis route is incompatible with 
	       the production of large amounts of $\alpha$~boron.

	       \begin{figure}[ht!]
		       \centering
		       \subfigure[]{\includegraphics[height=7cm]{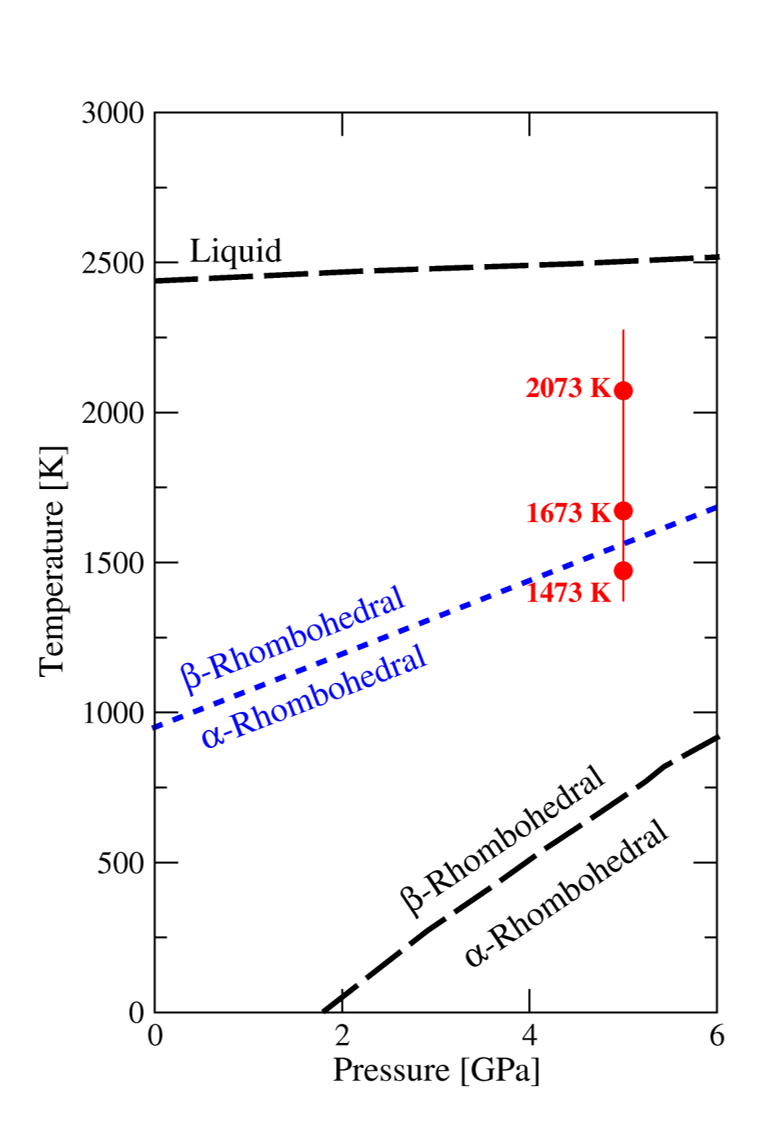} \label{fig:expt_phase_diagram}}
		       \subfigure[]{\includegraphics[height=7cm]{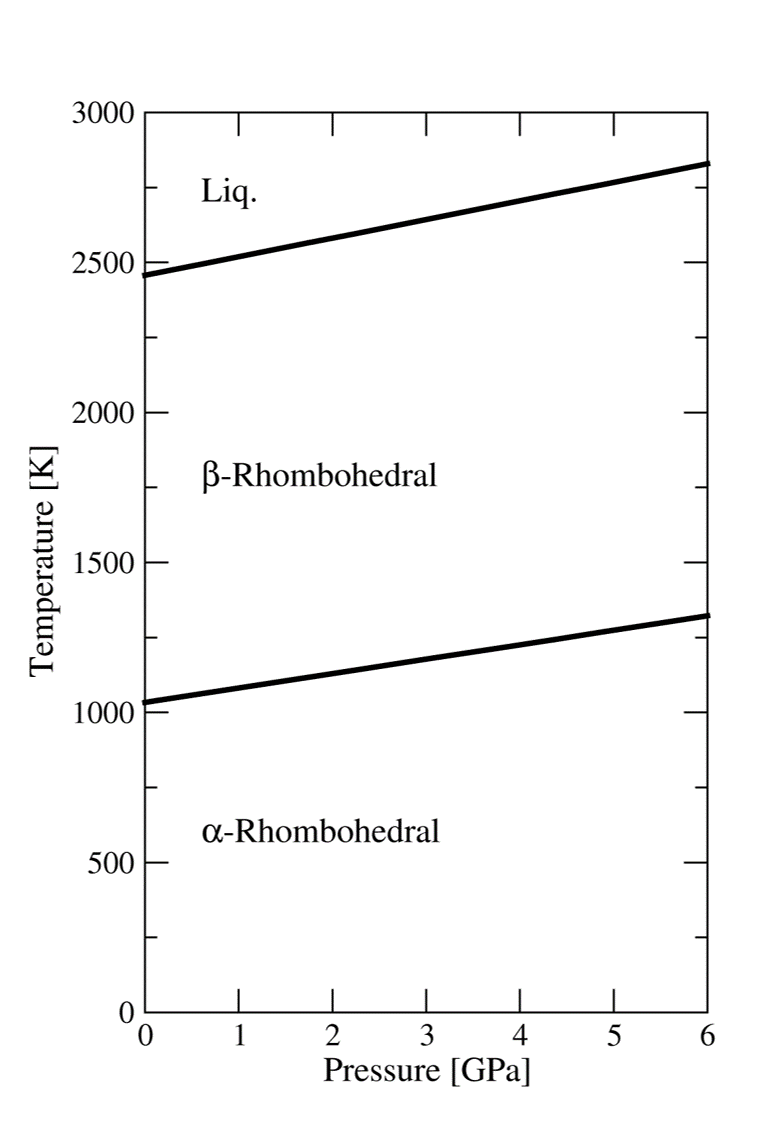} \label{fig:theo_phase_diagram}}
		       \caption{\label{fig:boron_phase_diagram} (P,T) phase diagrams of elemental boron. Right panel: theory~\cite{Shirai:2007}. Left panel: experiments. Black dashed line from Ref.~\cite{Oganov:2009} also cited by Ref.~\cite{Ogitsu:2013}. Blue dotted line from Ref.~\cite{Parakhonskiy:2011}. Red solid line: This work. Previous studies show consensus at 5~GPa and above 1570~K that the phase of boron is $\beta$~rhombohedral. However, in the present study, $\alpha$~boron has been synthesised in presence of carbon and remains as the dominant product (red disks) until 2073~K at 5~GPa. 
However, no $\alpha$~boron forms under the same experimental conditions when $\beta$~boron is used as the sole starting material, without carbon. 
}
	       \end{figure}

	       In the present work, we show experimentally that $\alpha$~boron forms at temperatures between 1473~K and 2273~K from crystalline~$\beta$ or amorphous boron under a pressure of 5~GPa (figure~\ref{fig:expt_phase_diagram}, red solid line) in presence of amorphous carbon, which is a more affordable route of synthesis than currently available methods. The syntheses were performed with the Paris-Edinburgh cell, samples were characterised with X-ray diffraction (XRD), Raman spectroscopy and optical properties -~their red colour~- and identified as $\alpha$~boron. The lattice parameters have been deduced from Rietveld refinement. 
	       This formation of $\alpha$~boron suggests that the presence of impurities is able to trigger the synthesis of $\alpha$ boron and explains differences in the various phase diagrams available in the literature (figure~\ref{fig:boron_phase_diagram}). 

Indeed, despite decades of studies~\cite{Shirai:2007,Oganov:2009,Parakhonskiy:2011,Shirai:2017}, there is no consensus over the boron phase diagram, especially at high temperatures and pressures.
The most stable phase at ambient conditions has been established to be $\beta$~rhombohedral boron, $\alpha$~boron being metastable~\cite{Oganov:2009,Shalamberidze:2000}. This seems reasonable considering that $\beta$~boron is widely available commercially, unlike $\alpha$~boron. Additionally, $\alpha$~boron is known to transform to $\beta$~boron at 950~K at ambient pressure, a signature of metastability~\cite{Amberger:1981}. The $\alpha$-to-$\beta$~phase transformation also occurs at any finite pressure (figure~\ref{fig:expt_phase_diagram}, black dashed line) up to 7.5~GPa, beyond which $\alpha$~boron transforms to $\gamma$~boron~\cite{Parakhonskiy:2011}. 
Instead of the conclusion of metastability, another interpretation of the phase diagram is that the phase boundary from the $\alpha$ to the $\beta$ phase should be defined as  
the limit of thermodynamical stability of $\alpha$~boron at high pressure and low temperature (figure~\ref{fig:expt_phase_diagram}, blue dotted line)~\cite{Parakhonskiy:2011}. 
Nonetheless, 
in all of the works~\cite{Shirai:2007,Oganov:2009,Parakhonskiy:2011}, there is consensus that $\beta$~boron is the most stable solid phase beyond 1500~K for all pressures ranging from ambient to 5~GPa. Contrastingly, in the present work, $\alpha$~boron is shown to form at 5~GPa in presence of carbon, at a temperature between 1473~K and 2273~K (figure~\ref{fig:expt_phase_diagram}, red~solid~line).

Concerning theoretical studies to establish the stability of various boron polymorphs, the extremely complicated and still debated structure of $\beta$~rhombohedral boron is a major roadblock. 
However, the common wisdom is that $\alpha$~boron is metastable with respect to $\beta$~boron at high temperature~\cite{Ogitsu:2009}, in the same manner as diamond is a metastable phase with respect to graphite~\cite{Tateyama:1996}. 
In some calculations within the density functional theory (DFT) in the local density approximation (LDA), $\alpha$~boron has been found to be marginally more stable than $\beta$~boron at zero temperature and zero pressure~\cite{Masago:2006}, while above 1000~K $\beta$~rhombohedral boron was the most stable phase for pressure values smaller than 50 GPa~\cite{Shirai:2007,Shirai:2017}. 
In another DFT work with the generalized gradient approximation (GGA), $\beta$~boron was modelled with a slightly different atomic structure~\cite{Van_Setten:2007}. 
It made $\beta$~boron the thermodynamically stable phase at any temperature when zero point energy was considered, while $\alpha$~boron remained as the metastable phase. Finally, in another series of theoretical papers, it has been found that the $\beta$ phase would be stabilized by a macroscopic amount of intrinsic defects that decrease the free energy by both increasing the entropy and reducing the internal energy~\cite{Ogitsu:2009}.

     It must be pointed out that phase changes of boron at high pressure and high temperature (HPHT) conditions reported so far in the literature have been observed either from pure elemental boron, or in some cases with the presence of metal catalysts. None of the studies took the effect of carbon 
on the formation of $\alpha$~boron into account, which is the purpose of the present work.
     In order to shed light on the role of carbon atoms, their hypothetical insertion into the atomic structure of $\alpha$~boron has been modeled within DFT with the aim of discriminating between two hypotheses: either carbon acts as a catalyst in the formation of $\alpha$~boron and the carbon atoms do not remain in the structure of $\alpha$~boron; or carbon acts as a dopant, and consequences of carbon insertion into the atomic structure on the physical properties need to be studied. 
				       In the present work, we have studied theoretically carbon insertion into the $\alpha$~boron atomic structure, and we have identified two kinds of low-energy defects, one of which is compatible with the volume variation and the red colour observed in our experiments, therefore tilting the scales towards the hypothesis that carbon acts as a doping atom.

       This paper is organised as follows~: In Sec.~\ref{sec:Methods}, both the experimental methods and the theoretical frameworks are presented. 
       Experimental results are reported in Sec.~\ref{sec:Expt_results}. Theoretical results on neutral and charged defects are reported in Sec.~\ref{sec:Theo_results}. 
       The combination of experimental and theoretical results is discusssed in Sec.~\ref{sec:combination} and conclusions are drawn in Sec.~\ref{sec:Conclusion}.

\section{Methods}
\label{sec:Methods}

\subsection{Materials and synthesis}
\label{sec:material_synthesis}

The methodology of the synthesis experiments performed in the Paris-Edinburgh press at 2 and 5~GPa has been discussed in earlier works~\cite{Chakraborti:2020,Chakraborti:2021}. In the present experiments, crystalline $\beta$~boron (Prolabo, 99.9\% purity, 1~to~10~$\mu$m grain size powder) or amorphous boron (Pavezyum, purity $>$98.5 \%, particle size $<$250~nm) was mixed with glassy amorphous carbon (Sigma-Aldrich, 99.95\% purity, powder grain of 2~to~20~$\mu$m) in a special non-commercial cBN mortar and pestle for 5~minutes in the stoichiometric ratio of 4:1. The large-volume two-column Paris-Edinburgh press VX5~\cite{Chakraborti:2021,Chakraborti:2020} 
was then used for the syntheses at two different pressures, 2~GPa and 5~GPa, under temperatures ranging from 1373~K to 2473~K.

Each sample has been compressed for approximately 4 hours, and once the target pressure was attained, the temperature has been increased at the rate of 2~K/s. The dwell time at the target temperature was 2 hours. The sample was then immediately quenched by turning off the power supply. In the next step, the sample has been decompressed over 6 hours. The gasket containing the sample has then been carefully broken to recover the synthesised product.

The temperature during the synthesis has been measured with a calibration curve that was obtained beforehand using the neutron radiography technique, which gives a precision of $\pm$20~K when under pressure. It had been verified at the synchrotrons SOLEIL (Gif-sur-Yvette, France) and ESRF (Grenoble, France) using the same power supply and cables as in presently reported \textit{ex situ} experiments. At the synchrotrons, the thermocouple had been inserted co-axially and the synchrotron beam used to measure the exact position of the thermocouple inside the sample (with radiography) -~these results were consistent with those from neutron radiography calibration curve~\cite{LeGodec:2004}.

The present pressure on the sample has been determined by the value of the primary pressure on the press, with a calibration curve obtained earlier through \textit{in situ} experiments at synchrotrons with pressure calibrants - the same methodology has been employed in earlier works \cite{Chakraborti:2020, Chakraborti:2021, Chakraborti:2022}. The uncertainty in the measured pressure value is $\pm$0.2~GPa.
The presence of $\alpha$~boron is easily identified by the red colour of the nanocrystal.
The synthesised product was characterised \textit{ex situ} with X-ray diffraction. The X-ray powder diffraction patterns have been recorded using a X'Pert ProPanalytical diffractometer equipped with a Cu~K$\alpha$ radiation source ($\lambda$~=~1.5406~\AA, detector: XCelerator) in the Bragg Brentano geometry (0.04\textdegree~Soller slits, 0.5\textdegree~programmable divergence slit, 1\textdegree~incident antiscatter slit, and 0.5\textdegree~diffracted antiscatter slit; filter iron). The process of the sample recovery from the gasket has sometimes led to the unwanted presence of a very small amount of boron nitride in the product, since boron nitride capsule formed the innermost part of the gasket assembly that holds the reactants~\cite{Chakraborti:2020}.

The Raman spectroscopy has been performed in the back-scattering geometry, using a Horiba Jobin Yvon HR800 Raman spectrometer with $\times$~10 and $\times$~20 objectives. The 514.5~nm line of an Ar$^{+}$ laser, with 5~$\mu$m beamspot, has been used to measure the atomic vibration frequencies in the samples.

The \textit{in situ} experiments were performed in the PSICHE beamline of the Soleil synchrotron. The samples were compressed to the
target pressure. The temperature was then slowly increased. At each
step of the temperature increase, a diffractrogram using energy dispersive X-ray diffraction (EDXRD) of the sample center
was produced using the white synchrotron beam. The
white beam has an energy range of 15–80 keV. It is focused to a size
of 25 μm in the vertical direction on the sample and collimated to 50 μm in
the horizontal direction. The diffractograms were taken at various
sample positions in order to ensure the homogeneity of the reactions. The EDXRD diffractograms were each taken for 2 minutes each. The data
obtained with the synchrotron radiation using a rotating solid state Ge detector has been converted to Cu K $\alpha$ in this work in
order to compare
the \textit{ in situ} data to the \textit{ex situ} results.

\subsection{Theoretical scheme} 
In the theoretical section, we have studied the formation enthalpy and volume changes of $\alpha$~boron in presence of carbon, in the form of various point defects, using the supercell approach. 

\subsubsection{Computational method}
Density functional theory (DFT) calculations of equilibrium volumes and total energies have been performed for supercells containing $N$~$\times$~$N$~$\times$~$N$ elementary unit cells of $\alpha$~boron with $N=2$~or~$3$ 
-~respectively $\sim$96 or $\sim$324 atoms~- using the plane wave and pseudopotential method, combined with the GGA in the PW91 functional (labeled DFT-GGA-PW91 in the following)~\cite{Perdew:1992}. The pseudopotentials for B and C atoms were those of Ref.~\cite{Jay:2019}. The size of the plane wave basis set was limited with a plane wave energy cutoff of 100~Ry.
The Brillouin zone (BZ) of the 2~$\times$~2~$\times$~2 (respectively, 3~$\times$~3~$\times$~3) supercells has been sampled by 4$^3$ (respectively, 3$^3$) Monkhorst-Pack (MP) \textbf{k}-point meshes~\cite{Monkhorst:1976}.  
       The neutral supercells for both pristine and defect-containing $\alpha$~boron, have been relaxed with respect to the cell shape, volume, and atomic positions in DFT-GGA-PW91. For charged supercells, the cell shape and volume were fixed to those of pristine and neutral $\alpha$~boron for the calculation of enthalpies~\cite{Bruneval:2015}, and to those of the neutral defect for the caculations of optical properties, while in both kinds of calculations the atomic positions were relaxed from those assigned in the neutral state. While the latter calculations have been performed with
a N = 2 supercell, the valence band maximum (VBM) and band gap of
pristine $\alpha$ boron have been computed with N = 3, from equations 3 and 4
given below.
To explore the effect of the carbon concentration on the volume (see section~\ref{sec:alpha_boron_discussion}), several defects have been inserted into the $N=2$ supercell. Defects were either put in proximity of one another - clustering defects- or at the maximal distances allowed by the supercell size.

The HSE06 hybrid exchange-correlation functional~\cite{Heyd:2003} with 25~\% of the exact exchange interaction has also been used to compute the Kohn-Sham eigenvalues and determine the band gap and energy level induced in the band gap by the defect. In that cases, $N=2$ supercells and a 4$^3$ \textbf{k}-point mesh passing through the $\Gamma$ point has been used.
For pristine $\alpha$~boron, HSE06 calculations have been performed 
with relaxation of the cell shape, volume, and atomic positions. For defective supercells, 
the relaxation of the atomic positions has been computed, while the unit cell volume and shape were conserved to that of pristine $\alpha$~boron. 
HSE06 calculations have been performed to overcome the drawback of DFT that tends to underestimate the band gap by $\sim$23\% in $\alpha$ boron~\cite{Cho:Note:2022:GGA_bandgaps}. 
Although formally, the GW approximation~\cite{Hedin:1965} has to be used, HSE06 was shown to provide reliable estimations of the band gap~\cite{Cho:Note:2022:xc_discontinuity}, as it turns out to be the case for pristine $\alpha$ boron also (see section~\ref{sec:Methods_optical_properties}). 

\subsubsection{Modelling of the defective atomic structures}
\label{sec:modeling_atomic_struc}
Various types of carbon defects have been studied: substitutional atoms, interstitial atoms, and complex defects. In the following, we will consider the two point defects that have the lowest formation enthalpy, which is sufficient for our purpose. The first configuration for carbon insertion in $\alpha$~boron is the substitutional carbon atom in the so-called \textit{polar} site of the icosahedron (figure~\ref{fig:boron_comparison}b). The second configuration is the interstitial defect where one carbon atom is located in the \textit{Td} site, with the tetrahedral symmetry~~\cite{Gunji:1993,Gunji:1996,Soga:2004,Dekura:2007,Dekura:2009,Dekura:2011,Hayami:2005}, which corresponds to one of the chain end positions in boron carbide (figure~\ref{fig:boron_comparison}a)~\cite{Dekura:2011}. Details on the other configurations for carbon insertion into the $\alpha$~boron atomic structure can be found in the supplemental material (SM, table S1). \\

\subsubsection{Evaluation of the formation enthalpy}
	To compare the formation enthalpy between systems that have different stoichiometries and charged states, we take the definition of the defect formation enthalpy as the following~\cite{Pantelides:1974,Pantelides:1974b}:
	\begin{equation}
		\label{eq:formation_enthalpy_main_equation}
		\Delta H_{D,q}(\mu_e,\mu_m)=(E_{D,q}-E_H) +\sum_{m=B,C} n_m \mu_m +q(E_V+\mu_e) + \frac{q^2\alpha_M}{2L\epsilon},
        \end{equation}
	\begin{equation}
		\label{eq:chem_pot_elements}
		\mu_m=\mu_m^0+\mu_m^*,
	\end{equation}

\begin{flushleft}	where \textit{D} and \textit{q} indicate the defect type and the charge of the defect (see also SM, table~S1). \textit{E$_{D,q}$} and \textit{E$_H$} are total energies of the defective system and the host material, which is in our case $\alpha$~boron. The difference between vibrational energies of the host and defective structures has been neglected. 
\end{flushleft}

The formation enthalpy is a function of \textit{$\mu_B$} and \textit{$\mu_C$}, the chemical potential of B and C atoms. 
$n_B$ and $n_C$ indicate the number of atoms exchanged with one reservoir of boron atoms and/or one reservoir of carbon atoms. They are positive when the atom is given from the system to one of the reservoirs and negative when the atom is supplied from one reservoir to the system.

$\mu_B^0$ and $\mu_C^0$ are the ideal chemical potentials per B or C atom, defined from the total energy values of the pristine $\alpha$~boron and diamond, respectively. $\mu_B^*$ is the excess chemical potential of a B atom that varies in our calculations from the boron-rich side ($\mu_B^*=0$) to the boron-poor side ($\mu_B^* \ll0$). For a C atom as well, the excess chemical potential $\mu_C^*$ varies in our calculations from the carbon-rich side ($\mu_C^*=0$) to the carbon-poor side ($\mu_C^* \ll0$)~\cite{Cho:Note:2022:excess_chem_pot_values,Raucoules:2011}.

	\paragraph{Charged defects} 

	Some defect configurations, and in particular the carbon atom substituted in the polar site of the icosahedron, (B$_{11}$C$^{\textit{p}}$), turn out to be metallic (SM, table~S1). 
	In order to make the system semiconducting, one additional electron has been subtracted ($q=+1$) or added ($q=-1$) to the system. For (B$_{11}$C$^{\textit{p}}$), we find that the energy is lowered w.r.t. the neutral case when the charged is positive.

The term $q(E_V+\mu_e)$ in Eq.~\ref{eq:formation_enthalpy_main_equation} accounts for such an exchange 
with a reservoir of electrons. 
The electronic chemical potential $\mu_e$ lies inside the band gap, \textit{i.e.} $0~\leq~\mu_e~\leq~E_g$, with respect to the valence band maximum (VBM). 
The energy of the VBM \textit{E$_V$} is defined as~\cite{Pantelides:1974,Pantelides:1974b,Dreizler:1990,Persson:2005}:

\begin{equation} 
               \label{eq:VBM} 
E_V=[E_{H,0}-E_{H,q}-E_M(L)]/q \Bigr|_{\substack{q=+1}},
\end{equation}
where the total energy difference between the neutral and defect-free system and the system in which one electron has been removed, is divided by the difference of charges between the two systems.
This equation will also be used to compute the band gap of the host material in Section~\ref{sec:Methods_optical_properties}. 

The last term in Eq.~\ref{eq:formation_enthalpy_main_equation}~and~\ref{eq:VBM} is the monopole term of the Makov-Payne correction, $E_M(L)=\frac{q^2\alpha_M}{2L\epsilon}$, which will be referred to as \textit{monopole correction} in the following, that compensates spurious interactions between the additional charge introduced in the supercell and its periodic images~\cite{Leslie:1985,Makov:1995,Lento:2002}. 
The resulting monopole correction amounts to 0.25~eV, for $\lvert q \rvert = 1$, which marginally modifies the formation enthalpy (SM, section S2,~\cite{Cho:Note:2022:Madelung}). 
This correction is accounted for in the following results, unless otherwise stated.

\begin{table}[ht!]
	\begin{threeparttable}
	\begin{tabular}{lllll}
	\hline \hline 
	& \multicolumn{2}{c}{$\alpha$~boron band gap}	& (B$_{11}$C$^{\textit{p}}$)	& (B$_{12}$)C$^{\textit{Td}}$	\\ 
	& Measured	          	& Theory (this work)  		&  $E_{def}$      		& $E_{def}$ 			\\ \hline 
	Expt.\tnote{a}		 			& 2			        &            			&				&					\\
		Expt.\tnote{b}                                  & 2.4                           &                               &                               &                                       \\
		Expt.\tnote{c}					& red colour\tnote{d}           &	                        &		                &			\\
		This work                                       & red colour\tnote{d}            & 				&				&                       \\
		$\Delta$SCF	{\small DFT-GGA-PW91}	        &                               & 2.15$\pm$0.25\tnote{e} 	& 1.89$\pm$0.08		        & 1.4$\pm$0.2				\\
		KS eigenvalues  {\small HSE06}                  &                               & 2.25$\pm$0.26\tnote{f}        & 1.96$\pm$0.21                 & 1.27$\pm$0.03          \\
		Defect level type        			&				&                               & $n$-type, shallow	        & deep		\\
		\hline \hline 
		\end{tabular}
		\begin{tablenotes}
		\footnotesize
		\item[a]{Optical absorption, from Ref.~\cite{Horn:1959}.}
		\item[b]{Electron energy loss spectroscopy (EELS), from Ref.~\cite{Terauchi:1997}.}
		\item[c]{From Ref.~\cite{Parakhonskiy:2011b}.}
		\item[d]{\textit{I.e.}, from 1.722~eV to 1.968~eV.}
		\item[e]{From relaxed positions of charged $\alpha$ boron~\cite{Cho:Note:2022:dscf_bandgap}.}
		\item[f]{From CBM at $F$ and VBM at $Z$.}
		\end{tablenotes}
		\caption{\label{tab:calcul_Eg}Theoretical band gap E$_g$ (eV) and defect-induced energy level 
			$E_{def}$~(eV) of pristine $\alpha$~boron and of defective supercells containing (B$_{11}$C$^{\textit{p}}$) and (B$_{12}$)C$^{\textit{Td}}$, obtained by the $\Delta$SCF method or with the KS eigenvalues 
				in HSE06. 
				The position in the band gap of the defect-induced energy level is given w.r.t. the VBM at the $\Gamma$ point of the supercell, where the $Z$ point of the elementary unit-cell is refolded. 
		}
\end{threeparttable}
\end{table}

\subsubsection{Optical properties}
\label{sec:Methods_optical_properties}
In the context of explaining the red colour of the samples, two computational methods have been used:
the $\Delta$SCF method, and Kohn-Sham (KS) eigenvalues in HSE06.  \\

\paragraph{The $\Delta$SCF method} 

The first band gap ($E_g$) has been computed as:\\
\begin{equation}
        \label{Eq:Eg}
        E_g=E_C-E_V,
\end{equation}
where $E_V$ is given by Eq.~\ref{eq:VBM} and $E_C$ is computed as:
\begin{equation}
        \label{Eq:CBM}
        E_C=[E_{H,0}-E_{H,q}-E_M(L)]/q \Bigr|_{\substack{q=-1}}.
\end{equation}
Total energies have been obtained within DFT-GGA-PW91 at the volume of neutral $\alpha$~boron, with relaxed atomic positions for the charged $N=2$ supercell. 

\paragraph{Kohn-Sham eigenvalues} 

Another method used to estimate the band gap of $\alpha$~boron and defect-induced energy levels is to compute the difference between the lowest unoccupied eigenvalue (KS LUMO) and the highest occupied one (KS HOMO). The KS eigenvalues $\varepsilon_{KS}$ were computed with the HSE06 hybrid functionals (table~\ref{tab:calcul_Eg}). 

As one can see from table~\ref{tab:calcul_Eg}, band gaps calculated from the HSE06 eigenvalues as well as from the $\Delta_{SCF}$ total energies are in good agreeement with experimental values. 
\paragraph{Defect-induced level} 

The energy level induced in the band gap have been identified by inspecting the HSE06 Kohn-Sham eigenvalue that were additional in the band gap, 
and accounting for an eventual refolding at the Brillouin zone center of the supercell. 

It has also been computed with the $\Delta$SCF method with the following equations: 
\begin{equation}
        \label{Eq:Edef_minus}
        E_{def}^-=[E_{def,0}-E_{def,q}-E_M(L)]/q \Bigr|_{\substack{q=-1}}, 
\end{equation}
when the level was occupied, such as for (B$_{11}$C$^{\textit{p}}$).
\begin{equation} 
               \label{Eq:Edef_plus}
        E_{def}^+=[E_{def,0}-E_{def,q}-E_M(L)]/q \Bigr|_{\substack{q=+1}},
\end{equation}
when the level was occupied, such as for (B$_{12}$)C$^{\textit{Td}}$.

Total energies have been obtained within DFT-GGA-PW91 at the volume of the neutral defect, with relaxed atomic positions for the charged $N=2$ supercell.

\section{Experimental results}
\label{sec:Expt_results}

\subsection{Synthesis and characterisation}

\begin{figure}[th!]
 \subfigure[] {\includegraphics[width=0.8\textwidth]{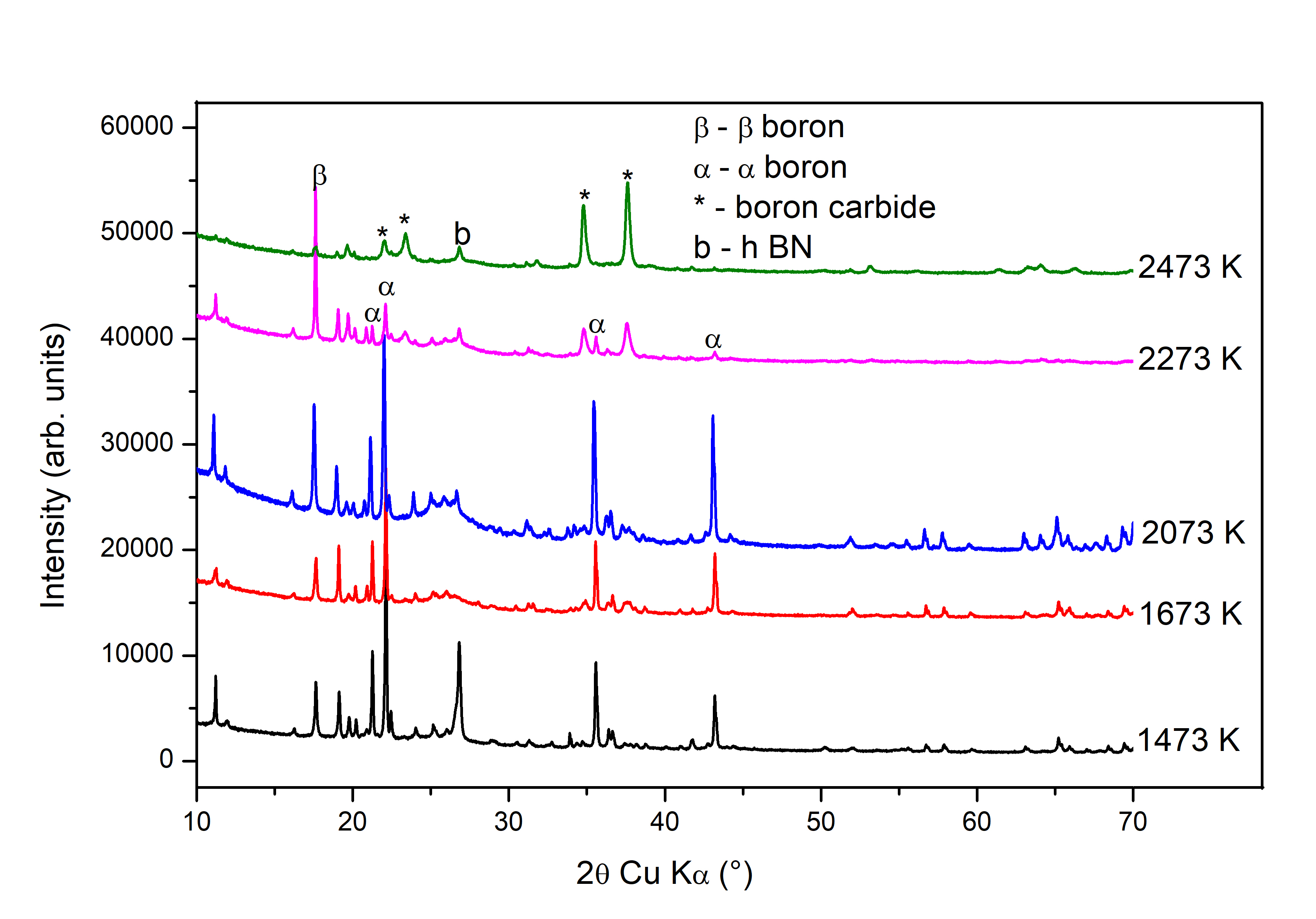} \label{fig:5GPa_B+C_full}}
 \subfigure[] {\includegraphics[width=0.8\textwidth]{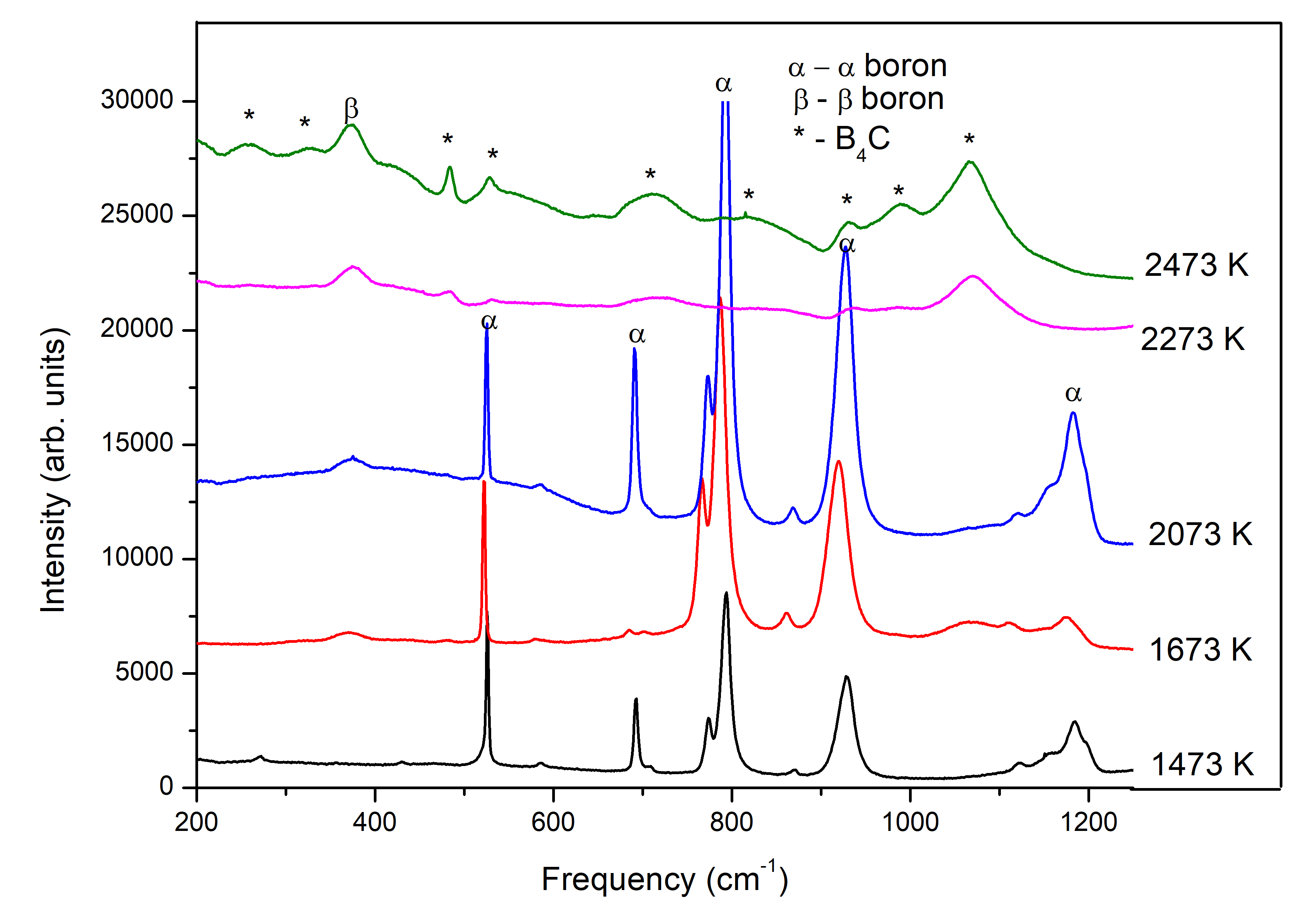} \label{fig:5GPa_B+C_Raman}}
        \caption{\label{fig:XRDpositions_5GPa_B+C} (a)~\textit{Ex situ} X-ray diffraction patterns taken at ambient P, T of quenched samples from mixtures of crystalline $\beta$~boron 
~\cite{Callmer:1977} and amorphous carbon under 5 GPa for 2 hours. Since $\beta$~boron 
has numerous peaks, only the most prominent ones have been marked, and all of the unmarked peaks are those of $\beta$~boron. $\alpha$~boron~\cite{Callmer:1977} 
is the most prominent phase until 2073 K. (b)~Raman spectra measured at ambient P, T of quenched samples from mixtures of crystalline $\beta$ boron and amorphous carbon under 5 GPa at various synthesis temperatures for 2 hours. The Raman peaks are consistent with the XRD results.
}
\end{figure}

 Figure~\ref{fig:5GPa_B+C_full} shows the X-ray diffraction patterns of samples from an initial mixture of crystalline $\beta$~boron and amorphous carbon quenched at ambient P,~T after a synthesis time of 2~hours at 5~GPa. $\alpha$ rhombohedral boron is the most prominent phase between 1473~K and 2073~K. This is corroborated by the Raman spectra shown in figure~\ref{fig:5GPa_B+C_Raman}.
 The size of the polycrystalline agglomerates formed during the synthesis was determined to be between 100~$\mu$m and 250~$\mu$m with optical microscopy.

\begin{figure}[b]
\vspace{-5cm}
\includegraphics[width=0.8\textwidth]{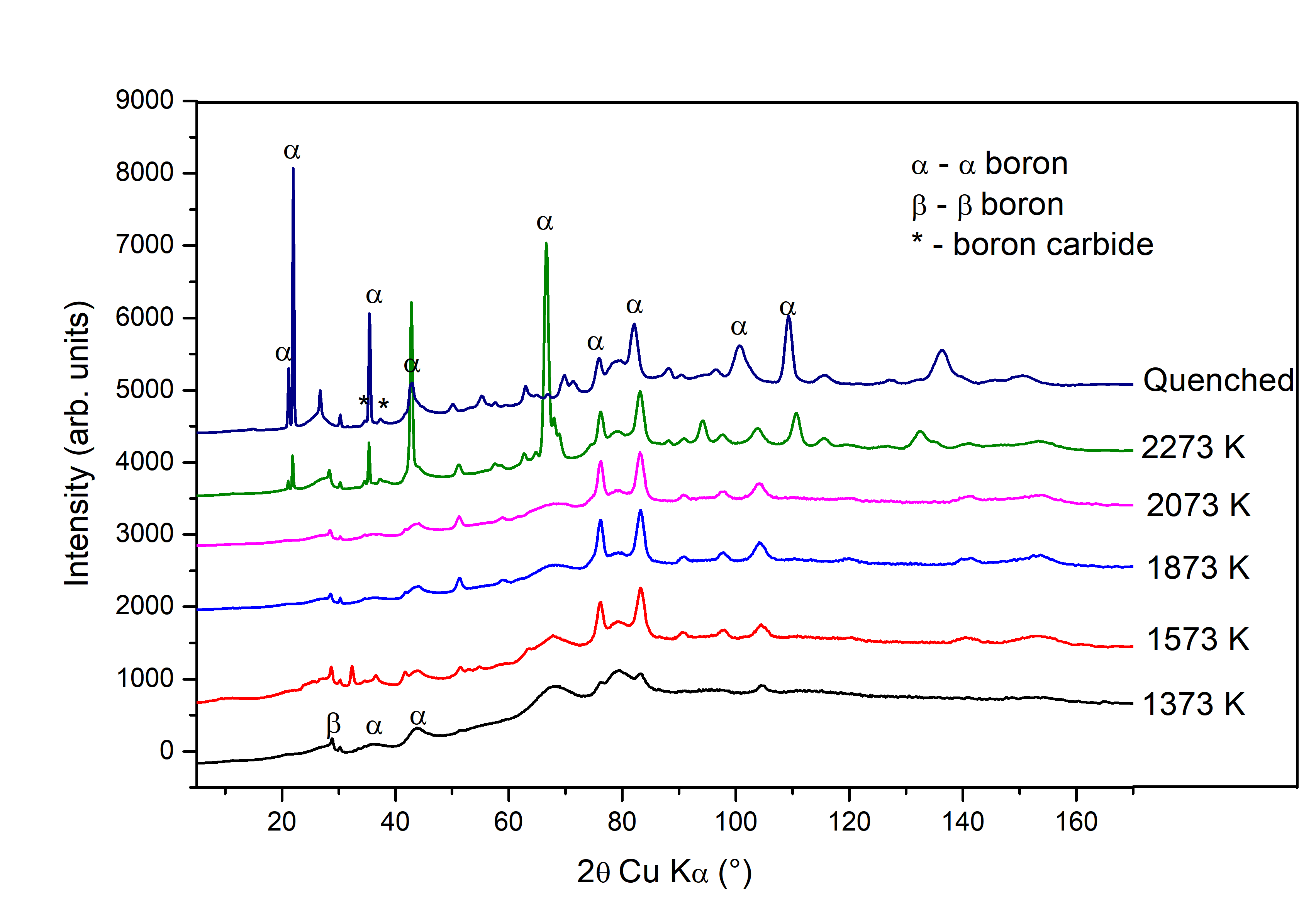}
\caption{\label{fig:5GPa_amorB+C_Raman} \textit{In situ} X-ray diffraction patterns obtained from mixtures of amorphous boron and amorphous carbon under 5 GPa at various temperatures. All of the unmarked peaks are those of $\beta$ boron~\cite{Callmer:1977}.
}
\end{figure}

		Similar results have been obtained \textit{in situ} at the Soleil synchrotron where amorphous boron and amorphous carbon were subjected to a pressure of 5~GPa (figure~\ref{fig:5GPa_amorB+C_Raman}), where changes have been evaluated in real time along with increasing temperature. $\alpha$~boron peaks are visible from 1373~K and remain until at least 2273~K. The boron carbide peaks appear clearly only after 
quenching from a synthesis temperature of 2273~K, a value slightly larger 
\FloatBarrier
\FloatBarrier
\noindent than when we used $\beta$~boron as a reactant instead of amorphous boron.
        However, we note that $\alpha$~boron formation from both \textit{i)} $\beta$~boron and amorphous carbon, and \textit{ii)} amorphous boron and amorphous carbon was not observed at the lower pressure value of 2 GPa~\cite{Chakraborti:2021, Chakraborti:2020}. 

\begin{figure}[t]
\begin{center}
 \subfigure[] {\includegraphics[width=0.7\textwidth]{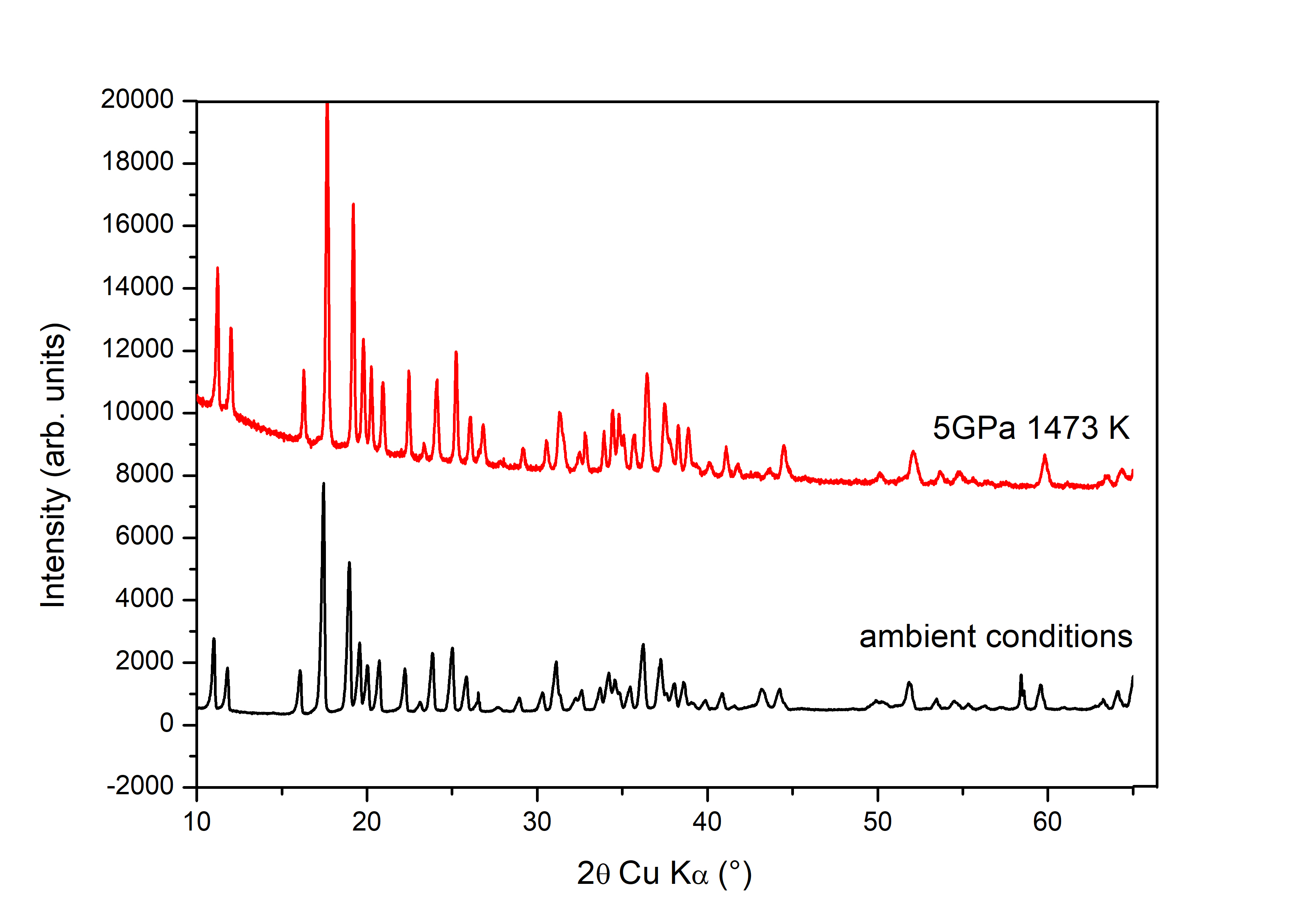} \label{fig:Bambient_B1200}}
 \subfigure[] {\includegraphics[width=0.65\textwidth]{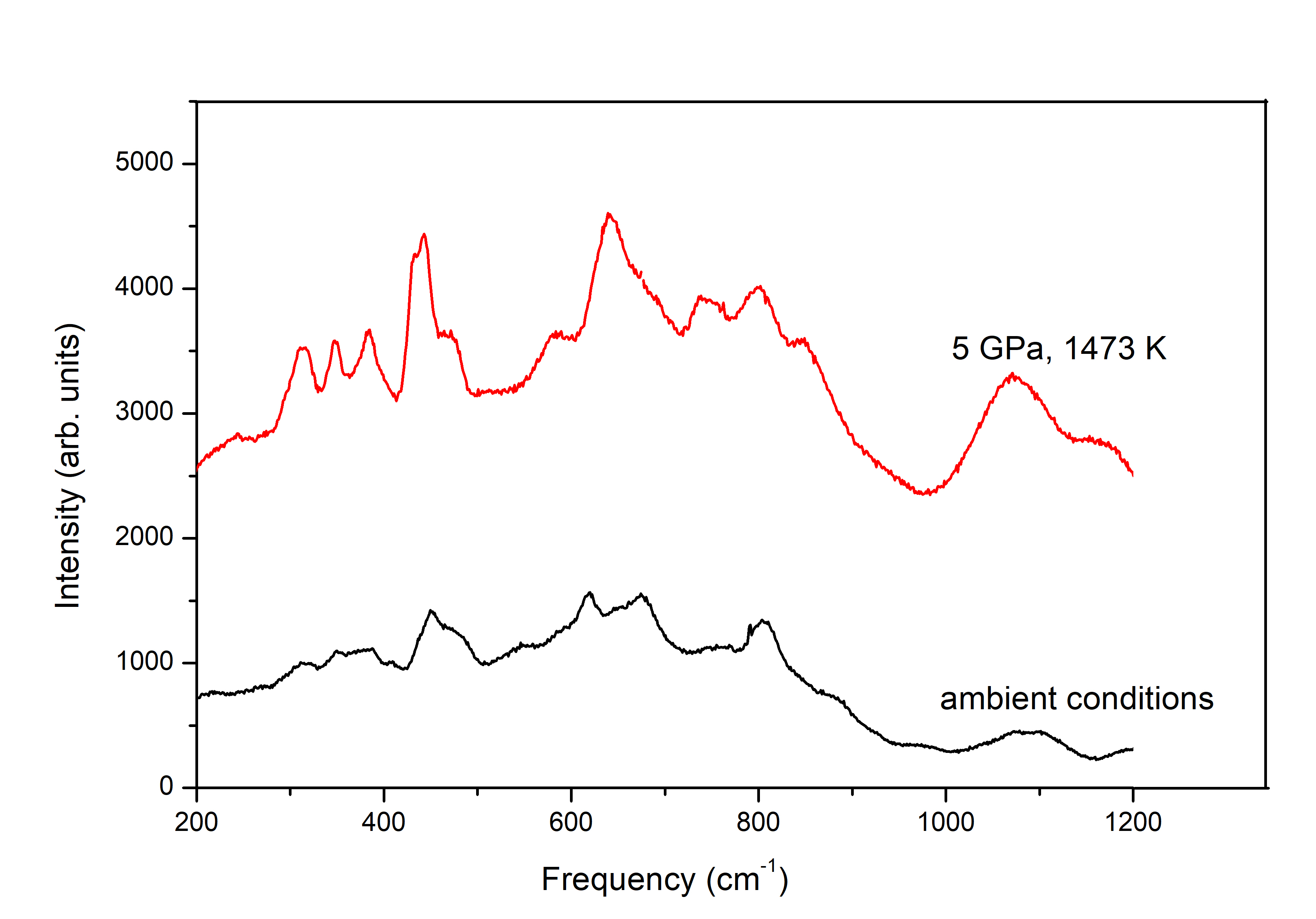} \label{fig:B_B1200_Raman}}
\end{center}
        \caption{\label{fig:XRDpositions_B} (a)~\textit{Ex situ} X-ray diffraction patterns measured at ambient P, T of the quenched sample of crystalline $\beta$~boron 
~\cite{Callmer:1977} subjected to 1473 K under 5 GPa for 2 hours, in comparison with that of $\beta$~boron not subjected to any HPHT treatment. All of the peaks are those of $\beta$~boron~\cite{Callmer:1977}. (b)~Raman spectrum taken at ambient P, T of quenched sample of crystalline $\beta$~boron subjected to 1473 K under 5 GPa for 2 hours, in comparison with that of $\beta$~boron not subjected to any HPHT treatment. 
}
\end{figure}

\begin{figure}[th!]
\centering
\includegraphics[width=0.6\textwidth]{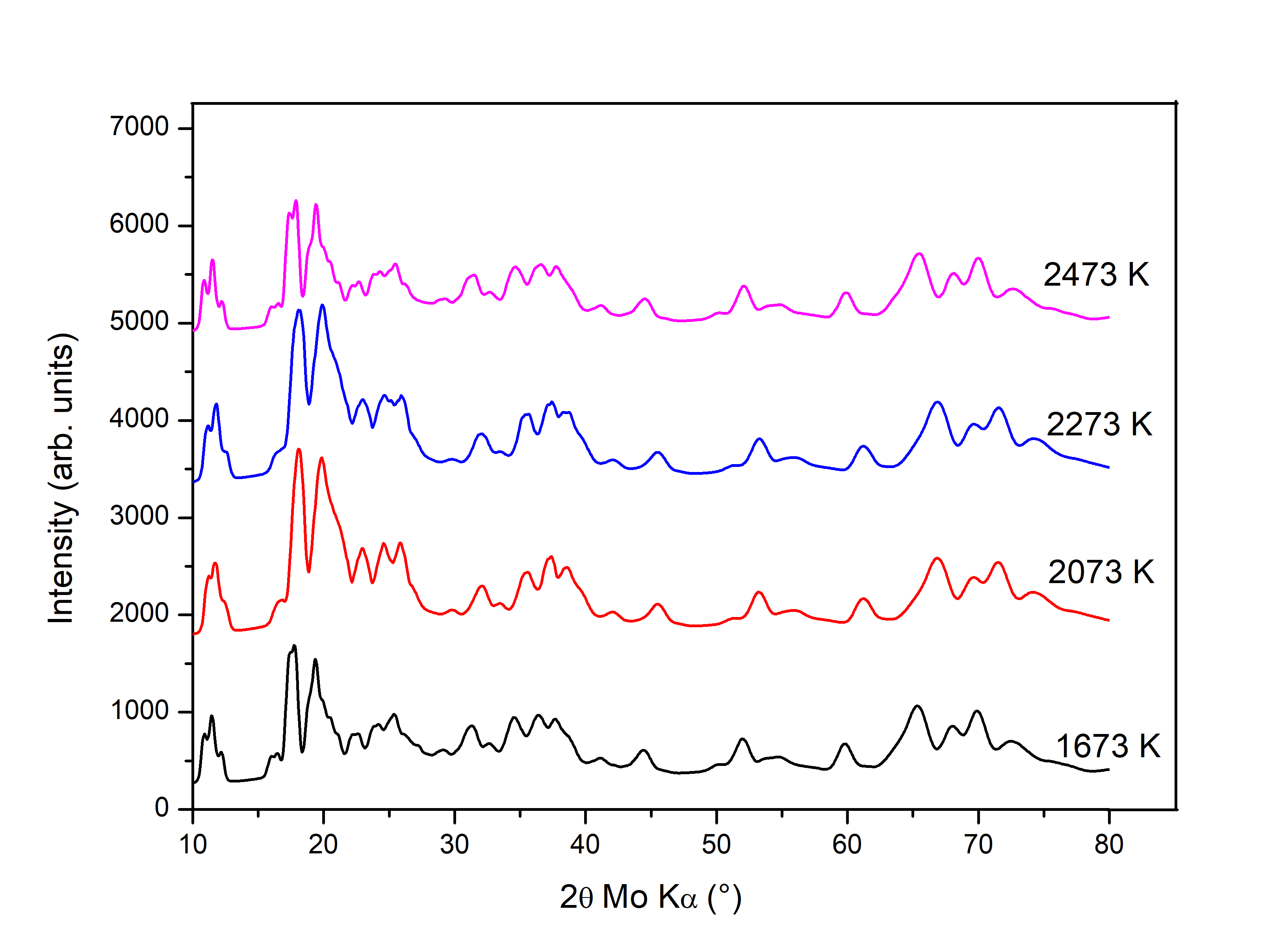}
        \caption{\label{fig:B_HPHT} \textit{Ex situ} X-ray diffraction patterns of quenched samples taken at ambient P, T of crystalline $\beta$~boron~\cite{Callmer:1977} subjected to various temperatures under 5 GPa for 2 hours. All of the peaks are those of $\beta$~boron~\cite{Callmer:1977}.}
\end{figure}

Similar experiments have then been performed on crystalline $\beta$~boron, solely without the addition of carbon, at various temperatures (1473~K, 1673~K, 2073~K, 2273~K, 2473~K, with a measurement error of $\pm$~20~K ) under a pressure of 5~$\pm$~0.2~GPa in the Paris-Edinburgh press, following the same methodology as above (figures~\ref{fig:XRDpositions_B} and~\ref{fig:B_HPHT}). The purpose of these experiments was to verify whether the amorphous carbon present in the reactant mixture had a role in the formation of $\alpha$~boron.

The first experiment in the series consisted of subjecting $\beta$~boron to a pressure of 5 GPa and a temperature of 1473 K for 2 hours. The quenched sample was then characterised at ambient P, T through XRD and Raman spectroscopy. Figures~\ref{fig:Bambient_B1200} and~\ref{fig:B_B1200_Raman} compare respectively the X-ray diffractograms and Raman spectra of the sample to those of crystalline $\beta$~boron that has not been subjected to any HPHT treatment.
The XRD pattern has shown no changes by HPHT treatment. The Raman spectrum of the HPHT-treated sample however shows sharper peaks and the peak positions can be measured more easily. This can be attributed to the recrystallization of $\beta$~boron under HPHT conditions, as previously documented~\cite{Parakhonskiy:2011,Solozhenko:2009}.

\FloatBarrier
\FloatBarrier

\FloatBarrier
\FloatBarrier

It is very important to note that $\alpha$~boron does not form when only $\beta$~boron is heated to 1473~K under 5~GPa. The $\beta$~boron reactant subjected to HPHT conditions
 above 1673~K were very hard and well-sintered, and therefore, they could not be ground into a fine powder in order to perform powder XRD with the copper anode. They were instead put inside a capillary and subjected to XRD using a Mo anode ($\lambda$ = 0.70931~\AA, figure~\ref{fig:B_HPHT}). $\alpha$~boron has not formed in any of these quenched samples in the absence of amorphous carbon.\\

\subsection{Discussion~: Presence of carbon impurities}
\begin{figure}[th!]
\begin{center}
 \subfigure[] {\includegraphics[width=0.7\textwidth]{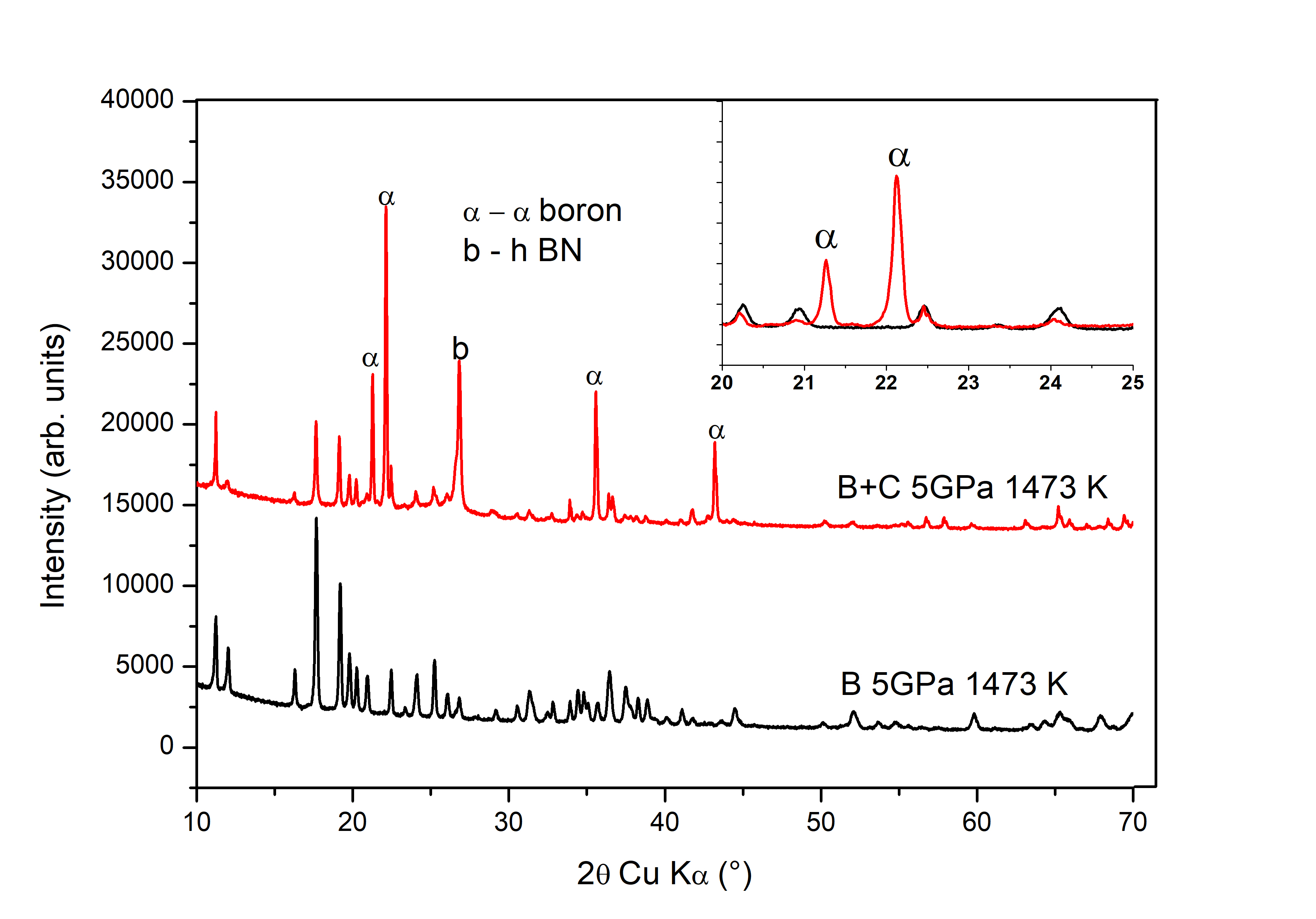} \label{fig:B1200_B+C1200}}
 \subfigure[] {\includegraphics[width=0.7\textwidth]{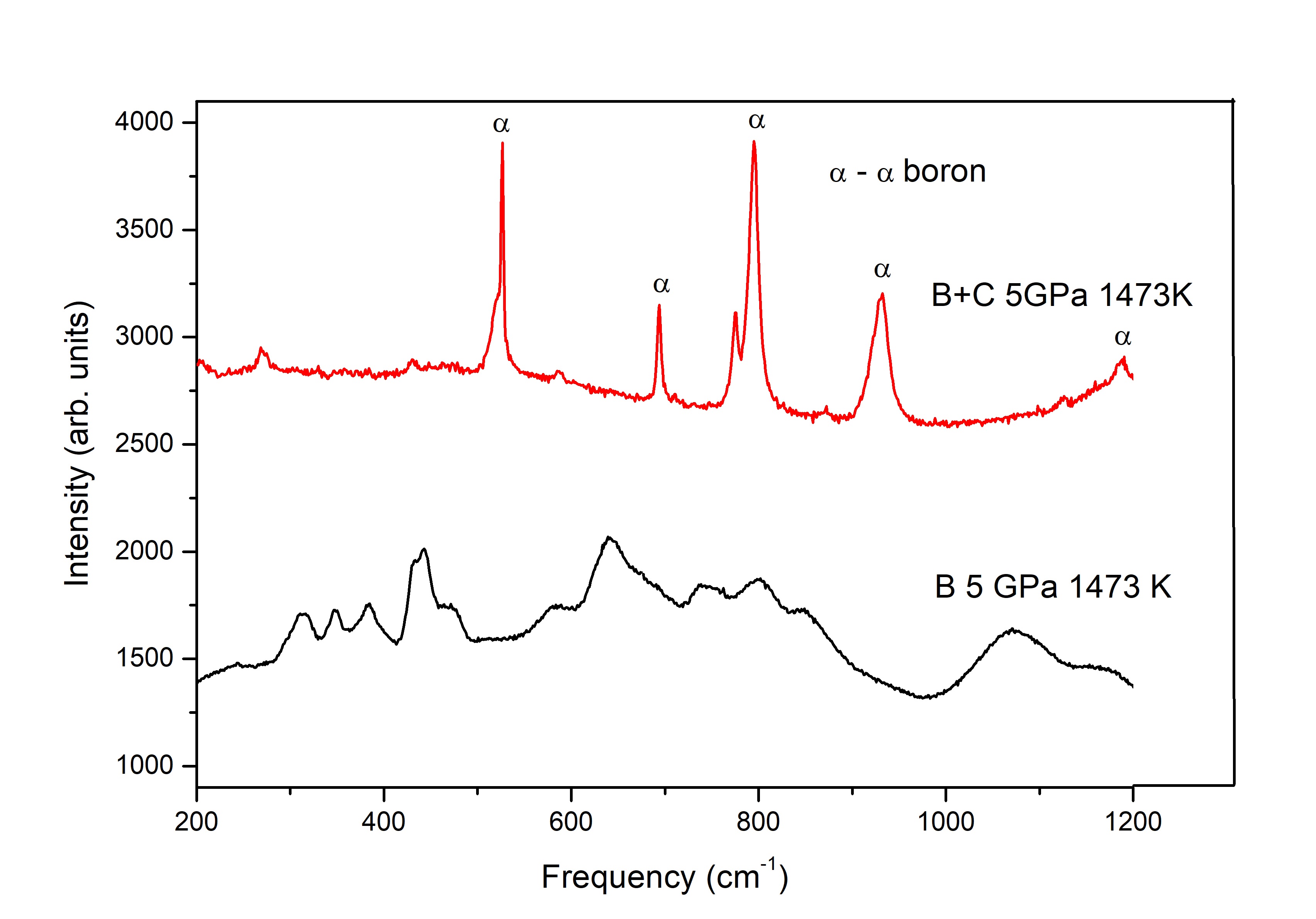} \label{fig:B1200_B+C1200_Raman}}
\end{center}
        \caption{\label{fig:XRDpositions_B+C} (a)~\textit{Ex situ} X-ray diffraction pattern measured at ambient P, T of quenched sample of a mixture of crystalline $\beta$~boron and amorphous carbon subjected to 1473~K under 5~GPa for 2~hours, in comparison with that of a quenched sample of $\beta$~boron subjected to the same HPHT treatment. All of the unmarked peaks are those of $\beta$~boron~\cite{Callmer:1977}. $\alpha$~boron~\cite{Callmer:1977} is formed in the quenched sample of the mixture of $\beta$~boron and amorphous carbon, unlike in the case when only $\beta$~boron was submitted to the HPHT treatment. (b)~\textit{Ex situ} Raman spectrum taken at ambient P, T of the quenched sample of a mixture of crystalline $\beta$~boron and amorphous carbon subjected to 1473~K under 5~GPa for 2~hours, in comparison with that of a quenched sample of $\beta$~boron subjected to the same HPHT treatment. All of the unmarked peaks are those of $\beta$~boron.}
\end{figure}

From the results described above, it is evident that carbon plays a role in the formation of $\alpha$~boron from $\beta$~boron at 5~GPa and temperatures up to 2273~K, and this is further highlighted by comparisons between the XRD patterns and Raman spectra of the products obtained with and without carbon (figure~\ref{fig:XRDpositions_B+C}). Therefore, it is worthwhile to verify whether carbon impurities are present in the atomic structure of the reactant, $\beta$~boron, or in the product, $\alpha$~boron. This is difficult to achieve experimentally, since carbon and boron atoms are both light elements with atomic numbers close to one another.
Thus, we deduced the presence or absence of carbon in the crystal structure by looking at the changes in the lattice parameters
of the quenched $\alpha$~and~$\beta$~boron at various temperatures of synthesis. To this end, we have performed the Rietveld refinement~\cite{Rietveld:2014} of the X-ray diffractograms using the FullProf program~\cite{Rodriguez_Carvajal:1990}.

\FloatBarrier
\FloatBarrier

        \begin{table}[th!]
                \centering
                \begin{threeparttable}
                        \scalebox{0.8}{
                        \begin{tabular}{lclllllllll}
\hline \hline
                                P$_S$ &        & \multicolumn{3}{c}{2 GPa}               & \multicolumn{3}{c}{5 GPa} & \multicolumn{3}{c}{Previous Expt. } \\
                                T$_S$ &{\small K}       & 1473          & 1673          & 2073   & 1473    & 1673    & 2073   &  \cite{Amberger:1981}      & \cite{Werheit:1993} & \cite{Callmer:1977}\\    \hline 
                                a$_h$ &{\small\AA }    & {\small 10.925 }       & {\small 10.922}        & {\small 10.929} & {\small 10.944}  & {\small 10.962}  & {\small 10.969} & {\small 10.944} & {\small 10.933} & {\small 10.9251 }\\
                                c$_h$ &{\small\AA }    & {\small23.807 }       & {\small 23.804 }       & {\small 23.807} & {\small 23.805}  & {\small 23.802}  & {\small 23.832} & {\small 23.8143} & {\small 23.8252} & {\small 23.8143 } \\
                                V  &{\small \AA$^3$/at.} & 7.81         & 7.81        & 7.82 & 7.84  & 7.86 & 7.88 & 7.84 & 7.83 & 7.81 \\
                                $\Delta$ & {\small \%}  & $-$0.38       & $-$0.44       & $-$0.30 & $-$0.04  & $+$0.28   & $+$0.53  &  & $-$0.16  & $-$0.35             \\ \hline \hline 
                        \end{tabular}}
                        \caption{\label{tab:Rietveld_beta_boron} $\beta$~boron after the HPHT treatment in presence of carbon atoms. Hexagonal lattice parameters and atomic volume, assuming 315~atoms per unit~cell~\cite{Cho:Note:2022:beta_structure}. 
The Rietveld analysis of XRD has been performed on the samples quenched at ambient P and T, the synthesis pressure and temperature being indicated by $P_S$ and $T_S$, respectively. The XRD diffractometer was calibrated using the (111) peak of a silicon (Si) wafer before each measurement, thus reducing instrumental error in the calculation of the lattice parameters. Previous experimental results for pristine $\beta$ boron are also provided for comparison~\cite{Amberger:1981,Werheit:1993,Callmer:1977}. $\Delta$ is the deviation of atomic volume with respect to the experiment of Ref.~\cite{Amberger:1981}. Comparison with table~\ref{tab:alpha_boron} confirms that the produced $\alpha$~boron is denser than $\beta$~boron~\protect\cite{Masago:2006}.}
                \end{threeparttable}
                \end{table}

\subsubsection{Lattice parameters of HPHT-treated $\beta$~boron} 
\label{sec:latt_para_beta_boron}
The lattice parameters of HPHT-treated $\beta$~boron in presence of carbon are consistent but not identical to those measured for pristine $\beta$~boron in previous works~\cite{Amberger:1981,Werheit:1993,Callmer:1977} (table~\ref{tab:Rietveld_beta_boron}). When $\beta$~boron is treated at 2~GPa, a volume contraction of the hexagonal unit cell that ranges between 2.1~\AA$^3$ and 3.2~\AA$^3$ has been obsserved, larger by factors of 4 to 6 than the volume variations with temperature that have been observed in pristine $\beta$~boron~\cite{Hiroto:2020}. 
The relative error on the atomic volume with respect to the experimental value of Ref.~\cite{Amberger:1981} lies between $-$0.30~\% and $-$0.44~\%. Such a decrease in volume with respect to pristine $\beta$~boron is consistent with the introduction of carbon into the atomic
structure, as measured in Ref.~\cite{Werheit:1993}.

\begin{figure}[ht!]
\includegraphics[width=0.9\textwidth]{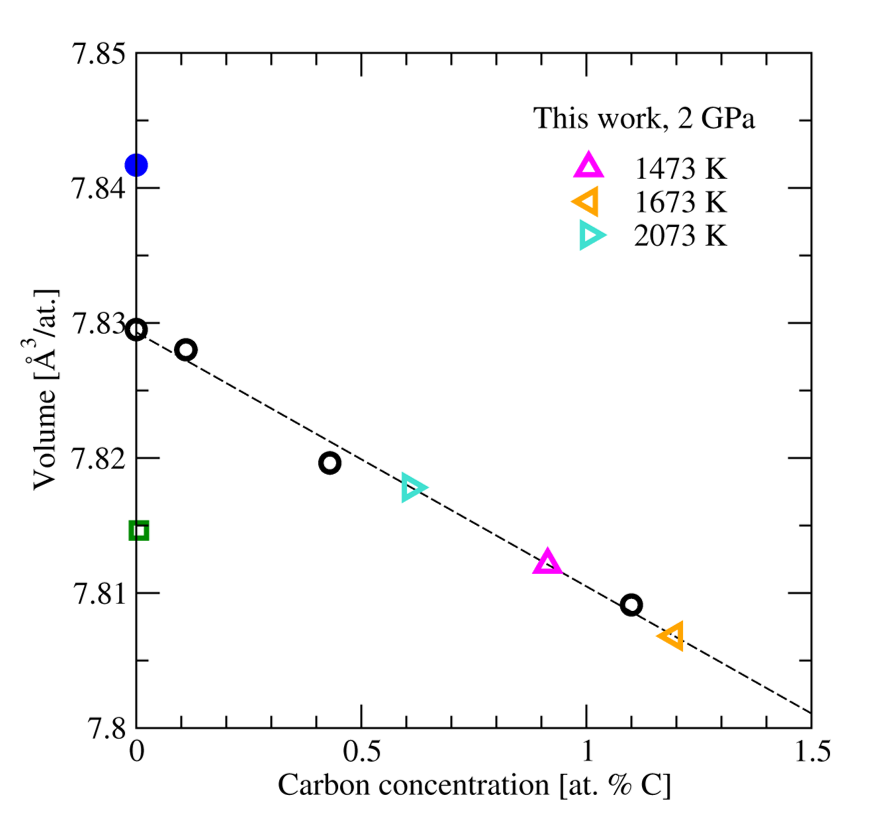}
        \caption{\label{fig:beta_boron_volume_2GPa} Atomic volume of $\beta$~boron HPHT-treated at 2 GPa in the present work (triangles), obtained from the Rietveld analysis in Table~\ref{tab:Rietveld_beta_boron}, compared with that of previous studies: black circles, green square, and blue disk from Refs.~\cite{Werheit:1993},~\cite{Callmer:1977}, and~\cite{Amberger:1981}, respectively. The black dotted line is a linear regression of the data from Ref.~\cite{Werheit:1993} supposing a hexagonal unit cell of $\beta$~boron containing 315 atoms~\cite{Cho:Note:2022:beta_structure}. The atomic volume values obtained in Table~\ref{tab:Rietveld_beta_boron} were then located on the regression line, and the carbon concentration in each of our samples has been deduced. It amounts to respectively 0.9~at.~\%~C at 1473~K, 1.2~at.~\%~C at 1673~K, and only 0.6~at.~\%~C at 2073~K, when boron carbide has formed.
        }
\end{figure}

Indeed, it has been experimentally shown in Ref.~\cite{Werheit:1993} that the substitution of carbon atoms in icosahedra of $\beta$~boron leads to a volume decrease. In particular, the authors assumed that the substitution took place in the polar site of the icosahedra in $\beta$~boron for geometric reasons. We use the volume change measured in Ref.~\cite{Werheit:1993} as a calibration curve (figure~\ref{fig:beta_boron_volume_2GPa}). It enables us to estimate the carbon concentration in our $\beta$~boron that has not transformed to boron carbide. The estimated carbon concentration lies between 0.6 and 1.2~at.~\%~C. It increases with temperature until 1673~K, but it decreases at 2073~K, when boron carbide has formed~\cite{Chakraborti:2021}.

In contrast, when $\beta$~boron is HPHT-treated at 5~GPa, the volume expands, with a change of the hexagonal u.c. volume varying from quasi-zero to 4.2~\AA$^3$, and a relative error on the atomic volume of $+$0.53~\% (table~\ref{tab:Rietveld_beta_boron}).
Moreover, it systematically increases with the synthesis temperature: the higher the synthesis temperature, the higher the volume measured at $\sim$~300~K.

        \begin{table}[ht!]
                        \centering
                        \begin{threeparttable}
                                \begin{tabular}{lllll}
                                \hline \hline 
                                  &    T$_S$~[K]& a$_h$ [\AA]                   & c$_h$ [\AA]                   & V [\AA$^3$/at.]                       \\ \hline
                This work          &    1473    & 4.904                         & 12.553                        & 7.262                                 \\
                                   &            & ( $-$0.08\tnote{a}~~)         & ( $-$0.11\tnote{a}~~)         & ( $-$0.27\tnote{a}~~)                 \\
                                   &            & ( $-$0.47\tnote{b}~~)         & ( $-$0.09\tnote{b}~~)         & ( $-$1.02\tnote{b}~~)                 \\
                                   &    1673    & 4.907                         & 12.563                        & 7.277                                 \\
                                   &            & ( $-$0.02\tnote{a}~~)         & ( $-$0.03\tnote{a}~~)         & ( $-$0.07\tnote{a}~~)                 \\
                                   &            & ( $-$0.41\tnote{b}~~)         & ( $-$0.01\tnote{b}~~)         & ( $-$0.82\tnote{b}~~)                 \\
                                   &    2073    & 4.912                         & 12.577                        & 7.300                                 \\
                                   &            & ( $+$0.08\tnote{a}~~)         & ( $+$0.08\tnote{a}~~)         & ( $+$0.25\tnote{a}~~)                 \\
                                   &            & ( $-$0.30\tnote{b}~~)         & ( $+$0.10\tnote{b}~~)         & ( $-$0.50\tnote{b}~~)                 \\
            Expt.\tnote{a}         & -          & 4.908                         & 12.567                        & 7.282                                 \\
            Expt.\tnote{b}         & -          & 4.927                         & 12.564                        & 7.337                                 \\
                                \hline \hline 
                                \end{tabular}
                                  \begin{tablenotes}
                                          \footnotesize
                                  \item[a]{With 0.04~wt\% of iodine and less than 0.003~\% of carbon~\cite{Decker:1959}.}
                                  \item[b]{From Ref.~\cite{Morosin:1986}.}
                        \end{tablenotes}
                        \end{threeparttable}
                        \caption{\label{tab:alpha_boron} $\alpha$~boron synthesised at 5 GPa. Hexagonal lattice parameters and atomic volume assuming 36~atoms per unit cell. The Rietveld analysis of the XRD has been performed on samples quenched at ambient P and T after the HPHT treatment in presence of carbon atoms. $T_S$ is the synthesis temperature. The deviation with respect to the experiment of Ref.~\cite{Decker:1959} and~\cite{Morosin:1986} are indicated in parentheses. No $\alpha$~boron has been found at 2~GPa.}
        \end{table}

	\subsubsection{Thermodynamical vs. kinetic processes}

        There is no formation of $\alpha$~boron at 2~GPa, and instead, boron carbide is formed at a temperature of synthesis (1873~K~\cite{Chakraborti:2021}) lower than that at 5~GPa, which is 2273~K. When syntheses are performed at 5~GPa, $\alpha$~boron is formed first, and boron carbide appears at a temperature of synthesis (2273~K) higher than that at 2~GPa, which is 1873~K~\cite{Chakraborti:2021}. Moreover, boron carbide is formed at the expense of $\alpha$~boron~\cite{Chakraborti:2020,Chakraborti:2021} in the sense that the higher the synthesis temperature, the smaller the quantity of $\alpha$~boron 
that is yielded.

From the above observations, we could hypothesise that at 5~GPa there is a kinetic competition between the formation of $\alpha$~boron and that of boron carbide.
Another interpretation could be that the formation of B$_4$C comes from the insertion of carbon into the atomic structure of $\alpha$~boron, resulting in a multi-step thermodynamical mechanism.
That -~as will be shown below~- carbon acts as a doping element and that the two low-energy sites of insertions are sites where carbon is present in boron carbide, make the scales tilt towards the interpretation
of a a multi-step thermodynamical mechanism. 

        \subsubsection{Lattice parameters of $\alpha$~boron formed at 5~GPa}
	\label{sec:latt_para_alpha_boron}

The formation of $\alpha$~boron from the mixture of $\beta$~boron and amorphous carbon requires an activation energy equivalent to a temperature of 1473~K under a pressure of 5~GPa.
The atomic volume (measured at 300~K) of $\alpha$~boron produced at HPHT ranges from 7.26~\AA$^3$/at.~to~7.30~\AA$^3$/at. (table~\ref{tab:alpha_boron}), when the synthesis temperature varies from 1473~K to 2073~K.
These volumes are in satisfactory agreement ($\pm$0.27~\%) with the previous report of Ref.~\cite{Decker:1959} which reported small amount of iodine impurities. However, our atomic volumes are systematically smaller than the more recent data of Ref.~\cite{Morosin:1986}. 

	Intriguingly enough, the atomic volume measured at 300~K varies with the synthesis temperature: the higher the synthesis temperature, the closer the lattice parameters from those of Ref.~\cite{Morosin:1986}.
	We therefore hypothesize the introduction of carbon atoms into the atomic structure of $\alpha$~boron yields a contraction of the volume. This hypothesis will further be substantiated with the calculations of Section~\ref{sec:Theo_results}.

	\section{Theoretical results}
	\label{sec:Theo_results}
\subsection{Study of the stability of neutral and charged defects}

\paragraph{Formation enthalpy of $\alpha$~boron with a neutral carbon defect} 
\label{sec:form_enthalpy_def_alpha_boron}

	The defect configurations that have the lowest formation enthalpies are more prone to be formed. 
        We find the lowest formation enthalpy (1.42~eV) in the case of the interstitial (B$_{12}$)C$^{\textit{Td}}$ defect.
        Then, the (B$_{11}$C$^{\textit{p}}$) defect has the second smallest formation enthalpy (1.76~eV). 
        The formation enthalpies for the other defects can be found in the SM (table~S1). 
 We conclude that insertion of carbon atoms into $\alpha$~boron turns out to be more favourable first in the \textit{Td} site, and second, in the \textit{polar} site of the icosahedra.

\begin{figure}[ht!]
	\centering
\subfigure[] {\includegraphics[height=6cm]{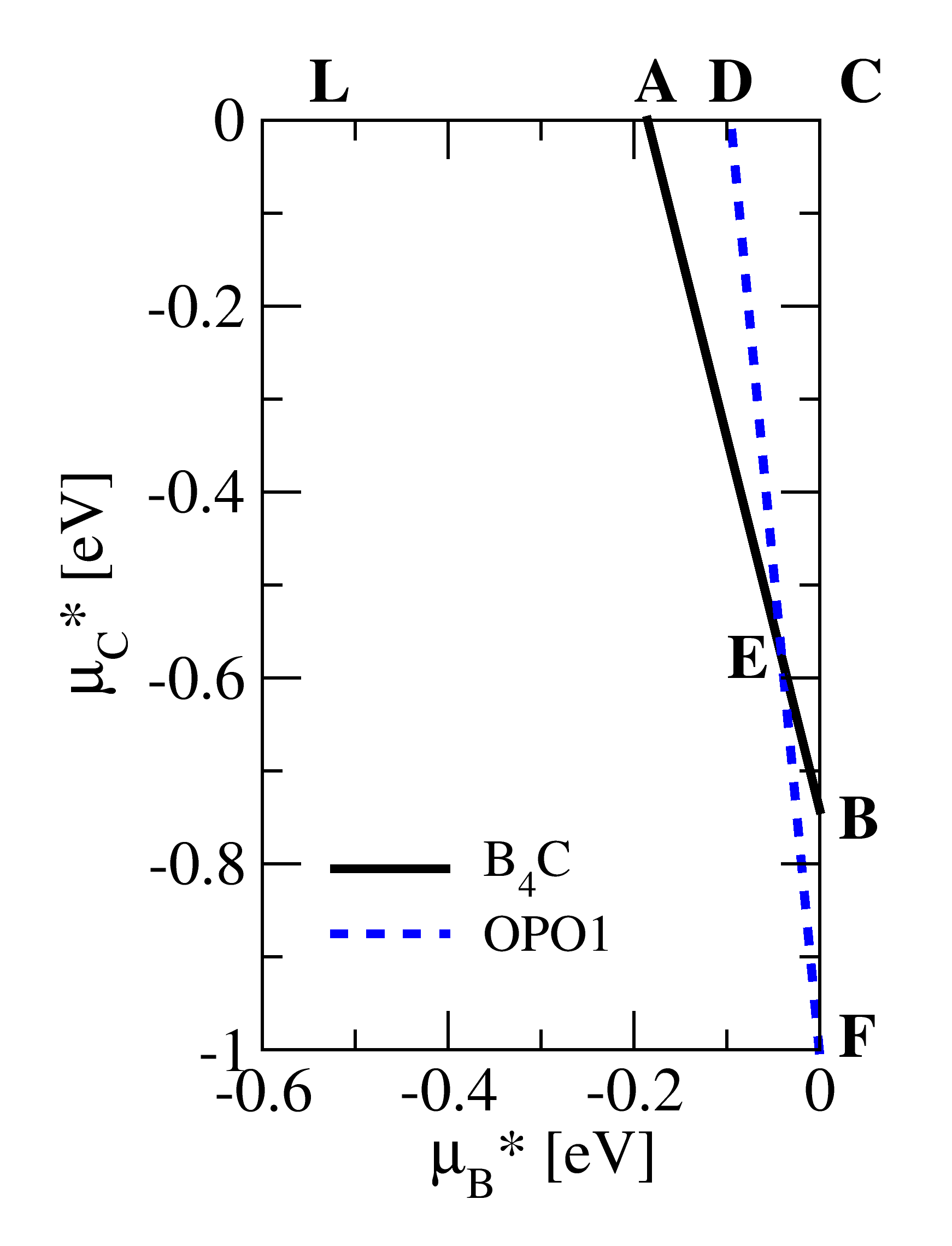} \label{fig:exc_chem_pot_B4C_OPO1}}
	\subfigure[] {\includegraphics[height=6cm]{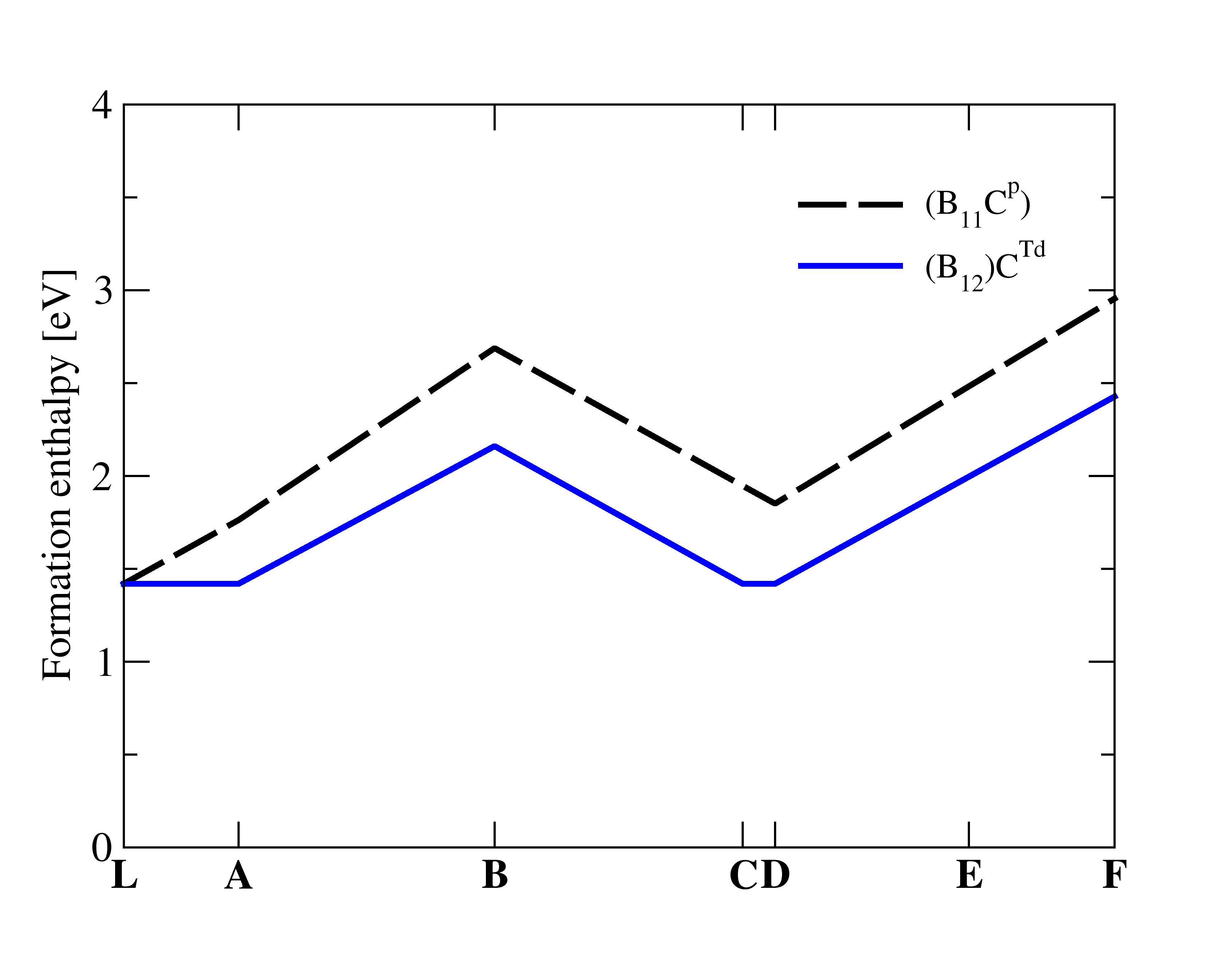} \label{fig:neutral_defects_compared}}
        \caption{\label{fig:neutral_defects_formation_enthalpy} (a) Plane defined by the values of the excess chemical potential of B and C atoms. 
The L, A, B, C, D, E, F points define the path along which the formation enthalpy has been computed in panel b. The black solid (respectively, blue dotted) line shows the relation between $\mu_B^*$ and $\mu_C^*$ defined by B$_4$C, boron carbide at 20~at.~\%~C (respectively, boron-rich boron carbide, OPO$_1$ at 8.7~at.~\%~C from Ref.~\cite{Jay:2019}). (b) The formation enthalpies of neutral defects along the \L--A--B--C--D--E--F path specified in (a). The (B$_{12}$)C$^{\textit{Td}}$ defect has the lowest formation enthalpy, and the (B$_{11}$C$^{\textit{p}}$) has the second one. At the point L, corresponding chemical potentials are $\mu_B^*=-$0.527~eV and $\mu_C^*=$0, where the formation enthalpies of (B$_{12}$)C$^{\textit{Td}}$ and (B$_{11}$C$^{\textit{p}}$) are equal.
        } 
\end{figure}

        In order to verify that this statement holds in general for any values of the chemical potentials, we vary the chemical potential of boron and/or carbon, and report the formation enthalpies of neutral (B$_{11}$C$^{\textit{p}}$) and (B$_{12}$)C$^{\textit{Td}}$ defects along particular directions in the ($\mu_B^*$,~$\mu_C^*$) plane (figure~\ref{fig:neutral_defects_formation_enthalpy}). Other defects have been studied in SM (section S3.1). In conclusion, the (B$_{12}$)C$^{\textit{Td}}$ is the most stable neutral defect for any chemical potentials of boron and carbon.

\paragraph{Formation enthalpies of charged defects in $\alpha$~boron} 

\begin{figure}[ht!]
                \includegraphics[width=0.9\textwidth]{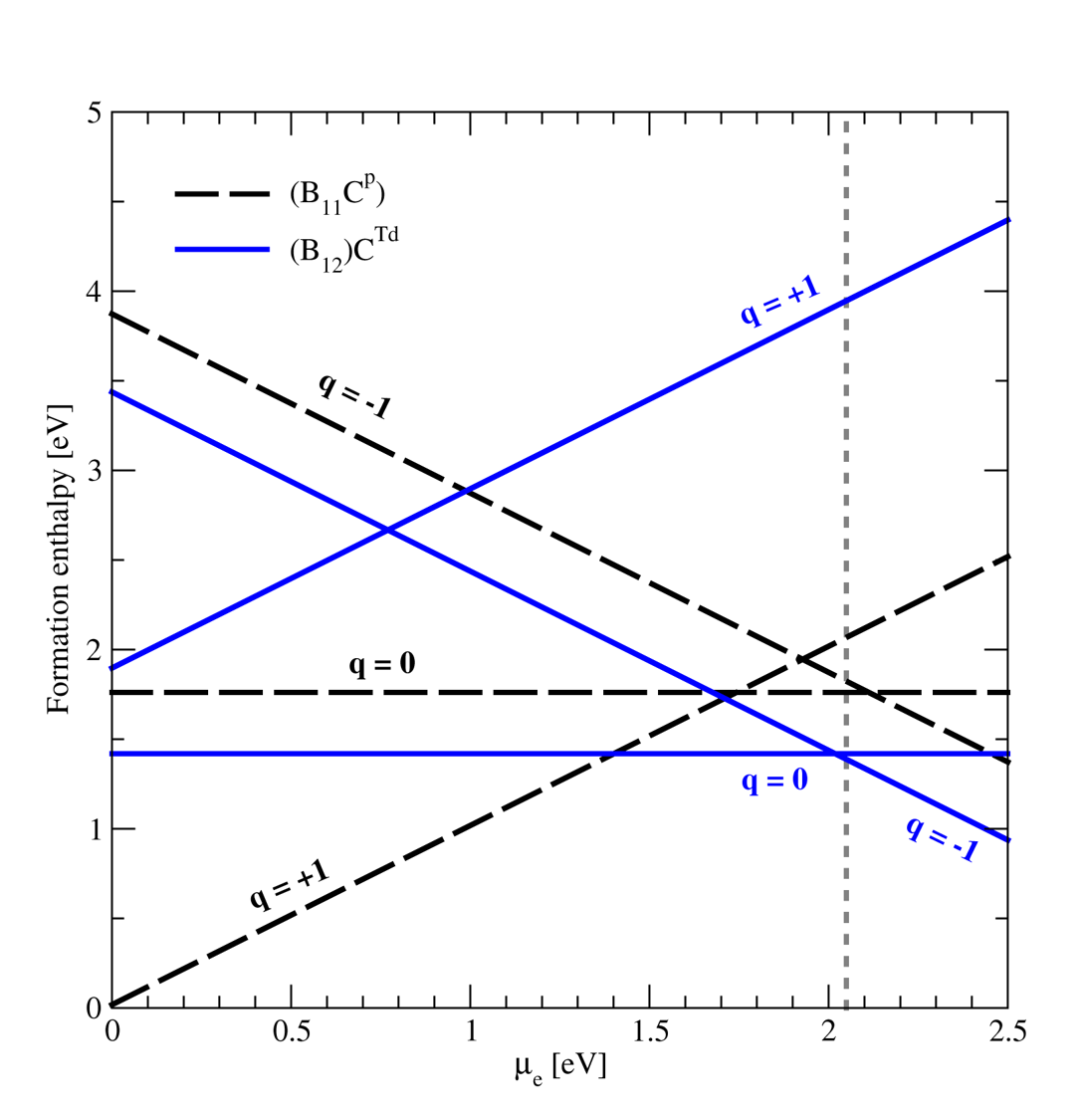}
                \caption{\label{sec:formation_enthalpy_wrt_el_chem_pot}
		Formation enthalpies of the (B$_{11}$C$^{\textit{p}}$) and (B$_{12}$)C$^{\textit{Td}}$ defects as a function of the electronic chemical potential (Eq.~\ref{eq:formation_enthalpy_main_equation}), with $\mu_C^*=0$ and $\mu_B^*=-0.19$~eV (point A in figure~\ref{fig:exc_chem_pot_B4C_OPO1}). A positive (respectively, negative or zero) slope indicates $q=+1$ (respectively, $q=-1$ or neutral) charge state.
A transition from the positively charged defect, (B$_{11}$C$^{\textit{p}}$)$^+$, to neutral and negatively charged (B$_{12}$)C$^{\textit{Td}}$ is seen at $\mu_e=$1.4 and 2~eV, respectively.
                $\mu_e=0$ corresponds to the VBM. The vertical lines indicates the band gap value obtained in DFT-GGA-PW91, by the $\Delta$SCF method (2.05~eV).  
                }
        \end{figure}

Remarkably, when positively charged, the (B$_{11}$C$^{\textit{p}}$) defect turns out to have a formation enthalpy smaller than that of (B$_{12}$)C$^{\textit{Td}}$ near the VBM of $\alpha$~boron (figure~\ref{sec:formation_enthalpy_wrt_el_chem_pot}, thick black solid line near $\mu_e=0$). The formation enthalpy as a function of the electronic chemical potential is reported in the SM (section 3.2) for defects other than (B$_{12}$)C$^{\textit{Td}}$ and (B$_{11}$C$^{\textit{p}}$).  
The very low formation enthalpy of (B$_{11}$C$^{\textit{p}}$) w.r.t. $\alpha$~boron can be understood by the fact the fact that the presence of the positive charge leads to a defect electronic configuration that has the same number of electrons as $\alpha$~boron. 

\subsection{Study of the volume of neutral defects}
	In the following, we are going to show that only the insertion of carbon atoms in the polar site of the icosahedra of $\alpha$~boron~-~(B$_{11}$C$^{\textit{p}}$)~- leads to a volume contraction and is in agreement with the observation of Sec.~\ref{sec:latt_para_alpha_boron}. The study of the equilibrium volume of pristine $\alpha$~boron both in HSE06 and in DFT-GGA-PW91 is detailed in the SM (section S4,~\cite{Vast:1997,Masago:2006}). The computational hurdle prevents the relaxation of the cell shape and volume in HSE06, so that the relative volumes w.r.t. pristine $\alpha$ boron have been computed in DFT-GGA-PW91 only. 

        \paragraph{Volume of $\alpha$~boron with one carbon defect} 

We stress that all of the atomic configurations of carbon based defects have been studied. 
	Interestingly, only in one case, when the carbon atom is substituted into the polar site of the icosahedron, does the volume show a decrease with respect to pristine $\alpha$~boron ($-$0.3~\%, SM, table~S1) which may help us identifying the (B$_{11}$C$^{\textit{p}}$) defect in experiments. Indeed, all of the other computed defect configurations lead to a volume increase (SM, table~S1). 
	For instance, for interstitial (respectively, complex) defects, the volume increase is larger than 1.2~\% (respectively, 0.7~\%) (SM, table~S1) whereas in the case of carbon substituted in the equatorial site of the icosahedron, the volume increase is small (0.3~\%).

\begin{figure}[th!]
\hspace*{1.6cm}
\includegraphics[width=0.9\textwidth]{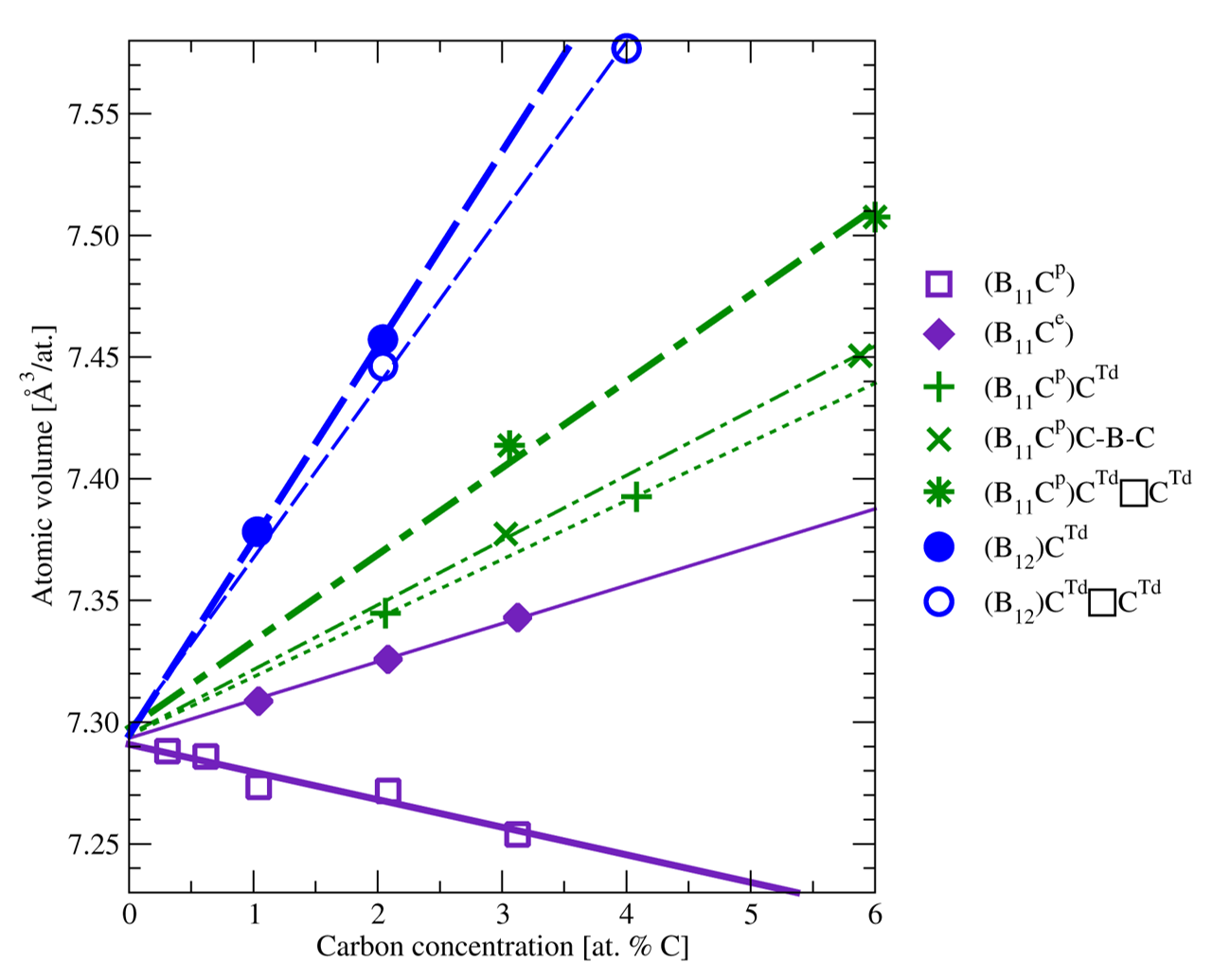}
	\caption{\label{fig:defective_volume_data} Neutral carbon based defects in $\alpha$~boron. Atomic volumes as a function of carbon concentration in DFT-GGA-PW91. The straight lines are linear regressions~\cite{Cho:Note:2022:slope_lin_reg_Vol}.
        }
\end{figure}
	\paragraph{Volume dependence on the carbon concentration} 
	\label{sec:alpha_boron_discussion}

	To explore the effect of carbon insertion on the volume further, we report the atomic volume as a function of carbon concentration (figure~\ref{fig:defective_volume_data}).
	A very important outcome of the present work is that we obtain a volume decrease only when carbon atoms substitute \textit{polar} boron atoms, (B$_{11}$C$^{\textit{p}}$), and this is true for any of the carbon concentration, both for clustering defects  (figure~\ref{fig:defective_volume_data}, squares and thick solid line) and for defects for larger distances between one another (not shown). In other words, unless there are interstitial atoms, carbon substitution in the polar site leads to a volume contraction.

Contrastingly, in all of the other cases, we find a volume expansion at any concentration (figure~\ref{fig:defective_volume_data}, symbols other than squares).
In each of these cases, the higher the number of defects, the larger the volume and none of the curves intersect. 
Thus, we propose the volume expansion or contraction as a fingerprint of a carbon defect type in $\alpha$~boron.

	From figure~\ref{fig:defective_volume_data} we also find that each defect can be characterised by the slope of the atomic volume as a function of the carbon concentration, 
	which may be helpful to interpret the experimental results.

\section{Combining theory and experiment}
\label{sec:combination}
	\subsection{Estimation of carbon concentration in our synthesised $\alpha$~boron} 
        \label{subsec:Exp_alpha_vs_calcul}
\begin{figure}[ht!]
                \centering
                \includegraphics[width=1.0\textwidth]{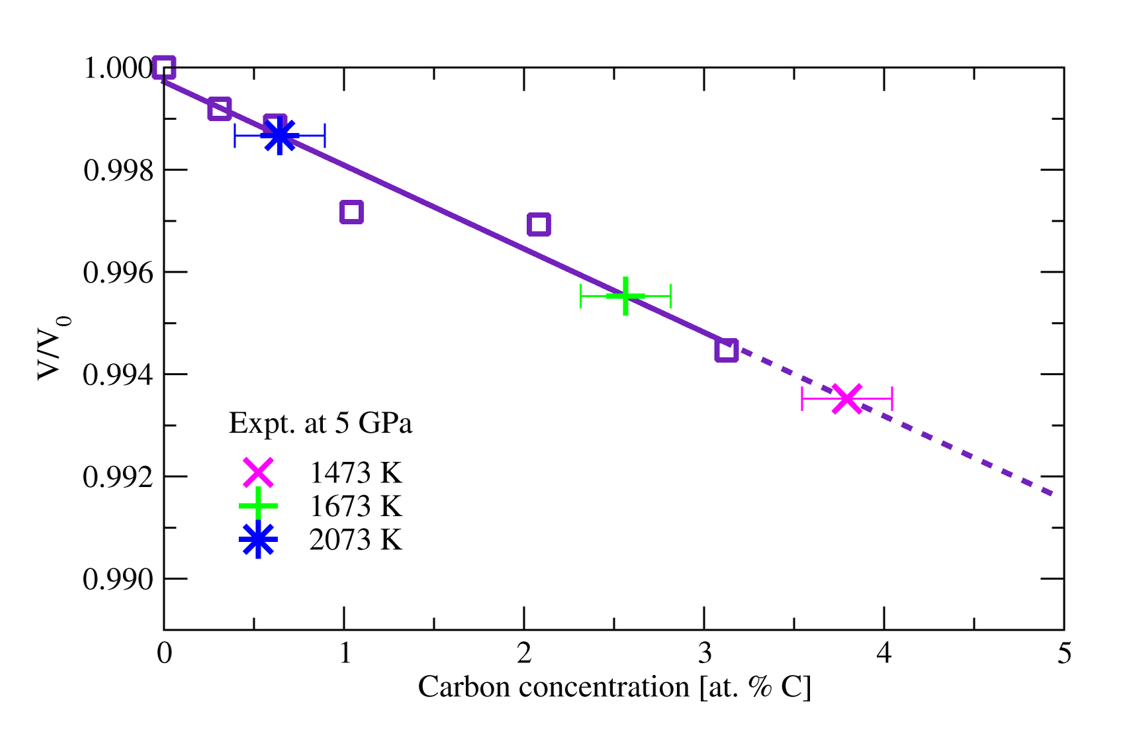}
		\caption{\label{fig:alpha_boron_5GPa} The ratio of our experimentally obtained $\alpha$~boron volume to the average experimental volume of pristine $\alpha$~boron from Refs.~\cite{Decker:1959,Morosin:1986}, projected onto the linear fit that comes from our calculations for (B$_{11}$C$^p$). The deduced carbon concentrations are 3.8, 2.6, and 0.6~at.~\%~C for the synthesis temperatures at 1473~K, 1673~K, and 2073~K. The violetline is the  linear regression of the DFT-GGA-PW91 calculations that has been done on the relative volume of the (B$_{11}$C$^{\textit{p}}$) defect as a function of carbon concentration~\cite{Cho:Note:2022:linear_fit_relative_Vol}, as done in figure~\ref{fig:defective_volume_data} for the volume itself.
}

        \end{figure}

	As the volumes of the experimentally obtained $\alpha$~boron samples are smaller than the one expected for pristine $\alpha$~boron~\cite{Morosin:1986}-~as reported in Section~\ref{sec:latt_para_alpha_boron}~- we expect carbon insertions in the form of \textit{polar} substitutions into $\alpha$~boron.
	We then deduce the carbon concentration of the experimentally obtained $\alpha$~boron from our calculations for the (B$_{11}$C$^{\textit{p}}$) defect (figure~\ref{fig:alpha_boron_5GPa}).
	The carbon concentrations are thereby estimated to be 3.8, 2.6, and 0.6~at.~\%~C as the synthesis temperature ranges from 1473~K to 1673~K, and to 2073~K, respectively with an error bar estimated to be~$\pm$0.25~\%. 
	We note that the higher the synthesis temperature, the lower the carbon concentration according to the volume comparison with our calculations.

	2073~K is interestingly  a temperature at which our sample volume is very close to that of pristine $\alpha$~boron.
$\alpha$ boron formed at temperatures higher than 2073~K is expected to have an even smaller carbon content than when formed at 2073~K, but the Rietveld refinement could not be performed at 2273~K -~for instance~- because of large widths of the XRD peaks. 
	In fact,  2073~K is the synthesis temperature limit where $\alpha$~boron is the dominant phase in the experiment: beyond this value, boron carbide is synthesised in the experiments, and less carbon atoms may be available for insertion into $\alpha$~boron.
	
Thus, synthesis temperature is a crucial parameter to control the carbon content during the production of $\alpha$ boron at 5 GPa.

        \begin{figure}[ht!]
                \includegraphics[width=1.0\textwidth]{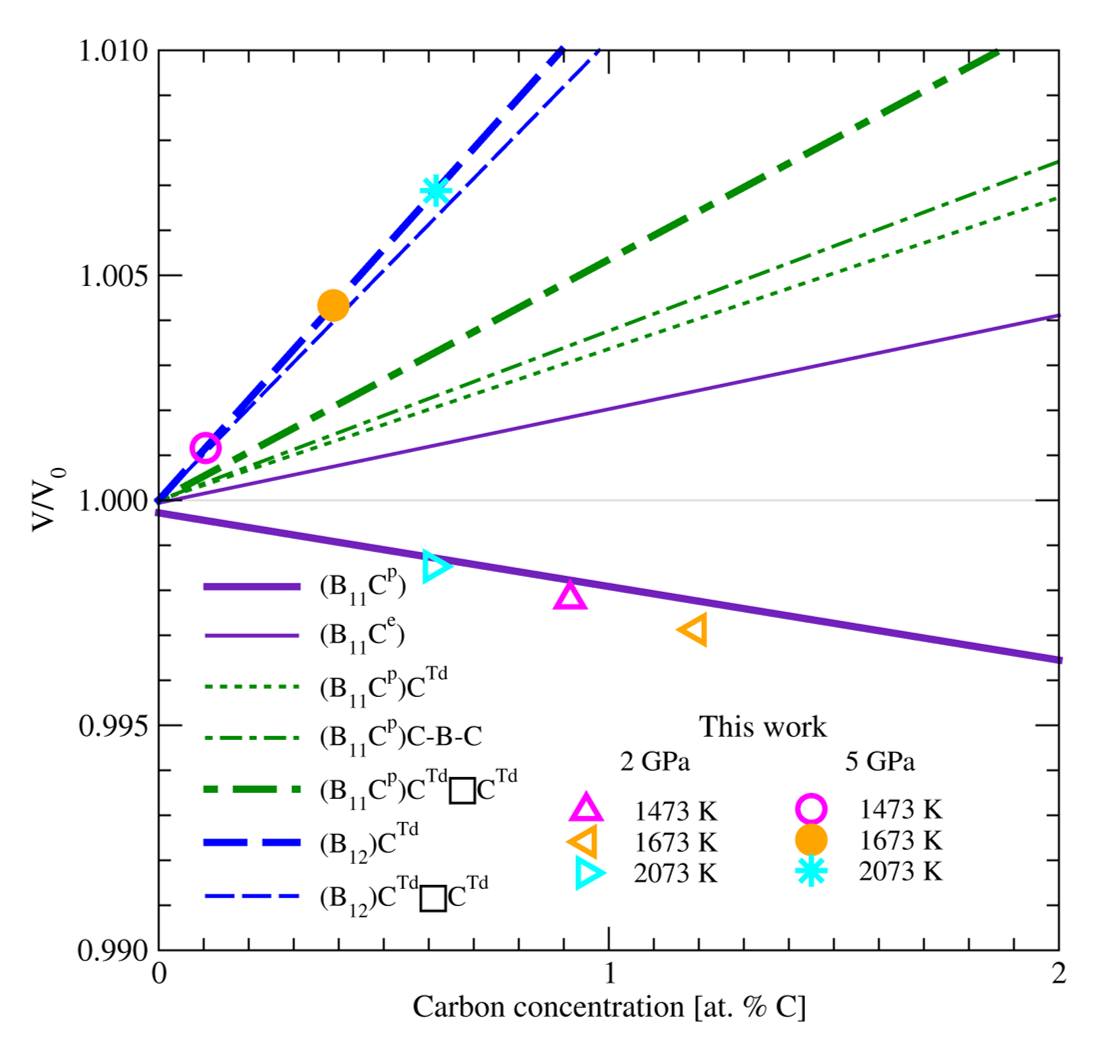}
                \caption{\label{fig:beta_boron_2GPa_5GPa} Ratio of the HPHT-treated $\beta$-boron volumes at 2~and~5~GPa, and of the volume of pristine $\beta$~boron from Ref.~\cite{Werheit:1993}, as a function of the carbon concentration deduced from the experiments from Ref.~\cite{Werheit:1993} (2~GPa) or projected onto the theoretical line on the (B$_{12}$)C$^{\textit{Td}}$ defect (5~GPa). The straight lines are the fit of the computed results of defective $\alpha$~boron presented in figure~\ref{fig:defective_volume_data}, changing $V$ to $V/V_0$. Colored dotted lines are the fits of the relative volume as a function of carbon concentration~\cite{Cho:Note:2022:slope_relative_Vol}.
                }
        \end{figure}

\subsection{Study of $\beta$~boron HPHT-treated at 2~GPa}
	\label{sec:Exp_beta_vs_calcul}
	Carbon insertion into $\beta$~boron has been so far studied as substitution in (B$_{12}$) icosahedra~\cite{Werheit:1993,Saengdeejing:2011}, which suggests a similar volume contraction in $\alpha$ and $\beta$~boron upon carbon substitution into (B$_{12}$) icosahedra.
	To further confirm the similarity, we now superimpose the experimental points defined by the carbon concentration (See Sec.~\ref{sec:latt_para_beta_boron}) and relative volume of $\beta$~boron HPHT-treated at 2~GPa onto those of computed defective $\alpha$~boron, using the ratio of our experimental volumes (table~\ref{tab:Rietveld_beta_boron}) and the volume of pristine $\beta$~boron from Ref.~\cite{Werheit:1993}.

	The behaviour of the relative experimental volume as a function of carbon concentration estimated from the calibration curve  (section~\ref{sec:latt_para_beta_boron}, figure~\ref{fig:beta_boron_volume_2GPa}) fits well with our calculations when the carbon atoms are substituted into the \textit{polar} site of the icosahedra of $\alpha$~boron (figure~\ref{fig:beta_boron_2GPa_5GPa}).
	This shows that, most probably, the carbon insertion into (B$_{12}$) icosahedra has similar chemical properties in $\alpha$ and $\beta$~boron, in terms of bond formation and electronic transfer. 

	Alternatively, we can apply the procedure of the last section (section~\ref{subsec:Exp_alpha_vs_calcul}) and project the experimental relative volumes onto the calculation for (B$_{11}$C$^{\textit{p}}$) in $\alpha$~boron.
	The estimated carbon concentrations turn out to be 1.0, 1.3, and 0.7~at.~\%~C at the synthesis temperatures of 1473, 1673, and 2073~K, respectively, in very close agreement -~$\pm$0.4\%~-with the carbon concentration obtained from the experimental result of Ref.~\cite{Werheit:1993} (figure~\ref{fig:beta_boron_volume_2GPa}).

\subsection{Discussion for $\beta$~boron HPHT-treated at 5~GPa}
        \label{sec:Exp_beta_5GPa_vs_calcul}
                In wide contrast to the previous section, at 5~GPa, the HPHT treatment of $\beta$~boron leads to a volume expansion.
                This indicates that carbon atoms are located at defect sites other than the \textit{polar} one.
                The comparison of $\beta$~boron experimental results with calculations on $\alpha$~boron must be taken with some caution. When we project the experimental relative volumes onto the curve of the least energetic defect with volume expansion, which is the (B$_{12}$)C$^{\textit{Td}}$ defect~\cite{Cho:Note:2022:def}, we find a carbon concentration smaller than 1~at.~\%~C in $\beta$~boron HPHT-treated at 5~GPa (figure~\ref{fig:beta_boron_2GPa_5GPa}, disk, circle, and star symbols). However,
        only a calculation of this defect configuration in $\beta$~boron would enable us to give a definitive conclusion on the nature of defects in $\beta$~boron HPHT-treated at 5~GPa.
        As the trigonal elementary unit cell of $\beta$~boron is not yet precisely known~\cite{Oganov:2009,Parakhonskiy:2011,Van_Setten:2007}, calculation of defective unit cells of $\beta$~boron is beyond the scope of the present work.

        \begin{table}[th!]
                \centering
                \begin{threeparttable}
                \begin{tabular}{llrlrll}
                       \hline \hline 
  :               &               & \multicolumn{2}{c}{$\alpha$~boron}    & \multicolumn{3}{c}{$\beta$~boron}     \\
       P$_S$ [GPa]& T$_S$ [K]     & site                  & at.~\%~C      & site                  & at.~\%~C      & at.~\%~C      \\
       2          &               &                       &               &                       &               &               \\
		  &1473           &                       &               & \textit{polar}        & 1.2           & 0.9           \\
		  &1673           &                       &               & \textit{polar}        & 1.6           & 1.2           \\
                  &2073           &                       &               & \textit{polar}        & 0.7           & 0.6           \\
       5          &               &                       &               &                       &               &               \\
		  &1473           & \textit{polar}        & 3.8           & \textit{interstitial} & 0.1           &               \\
		  &1673           & \textit{polar}        & 2.6           & \textit{interstitial} & 0.4           &               \\
	 	  &2073           & \textit{polar}        & 0.6           & \textit{interstitial} & 0.6           &               \\
                  &2273           &                       & $<$ 0.6       &                       &               &       \\
                          \hline \hline 
                \end{tabular}
                \end{threeparttable}
                \caption{\label{tab:graphs_data}Carbon concentrations and defect sites in $\alpha$~boron produced at 5~GPa and HT, and in $\beta$~boron HPHT-treated at 2 and 5~GPa. They are deduced from calculations of defective $\alpha$~boron (columns 3,~4,~5,~and~6), using figure~\ref{fig:alpha_boron_5GPa} for $\alpha$~boron at 5~GPa and figure~\ref{fig:beta_boron_2GPa_5GPa} for $\beta$~boron. The last column is the carbon concentration estimated with the experiments of Ref.~\cite{Werheit:1993} (figure~\ref{fig:beta_boron_volume_2GPa}).
                }
        \end{table}

\subsection{Optical properties of defective $\alpha$~boron}
        The (B$_{11}$C$^{\textit{p}}$) and (B$_{12}$)C$^{\textit{Td}}$ defects have widely different properties. While (B$_{11}$C$^{\textit{p}}$) is a shallow level defect close to the conduction band minimum leading to n-type doping, (B$_{12}$)C$^{\textit{Td}}$ induces a deep level defect (table~\ref{tab:calcul_Eg}).
Hence, we expect (B$_{11}$C$^{\textit{p}}$) to maintain the colour of $\alpha$~boron, while the presence of a deep level state for (B$_{12}$)C$^{\textit{Td}}$ is expected to give a black aspect to $\alpha$~boron.\\

\subsection{Determination of the sample carbon contents}
\label{sec:sample_carb_content}
The finding that the volumes of both the reactant and the product vary after HPHT treatment with respect to those of pristine boron, and evolve when the reaction temperature and/or pressure change, is a strong signature of carbon having been inserted into boron. The estimated carbon concentrations and insertion atomic sites are summarised in table~\ref{tab:graphs_data}.

Our examination of $\alpha$ and $\beta$~boron volumes in our present experiments and DFT calculations tends to show that carbon atoms act as a doping element rather than as a catalyst in the formation of $\alpha$~boron at 5~GPa, mostly in the form of (B$_{11}$C$^p$) defects. Moreover, we find that the occurrence of (B$_{11}$C$^{\textit{p}}$) defects marginally changes the colour of $\alpha$~boron, which we used as the signature of the formation of $\alpha$~boron in our experiments.    

The carbon concentration in $\alpha$ boron turns out to reach a few percent at low temperature, and to decrease by one order of magnitude when the synthesis temperature is increased (table~\ref{tab:graphs_data}). 

We also propose that carbon atoms stabilise the $\beta$~boron structure at 2~GPa before the formation of boron carbide. However, the carbon concentration in $\beta$~boron is 2~or~3 times smaller than that in $\alpha$~boron at 1473 and 1673~K. 

Finally, carbon insertion in $\alpha$ boron occurs in atomic sites where carbon is also present in boron carbide, favouring the interpretation that the formation of boron carbide 
is a multi-step thermodynamical mechanism in our experiments.

\section{Conclusions}
\label{sec:Conclusion}

The formation of $\alpha$~boron from a reactant mixture of $\beta$~boron and amorphous carbon, or from amorphous boron and amorphous carbon,  under a pressure of 5 GPa within a temperature range of 1473 K - 2273 K has been reported for the first time.

We show that the presence of carbon plays a vital role in the $\alpha$~boron synthesis, and that under the condition of carbon presence, a revised phase diagram of boron should be proposed.

The comparison of the experimental volumes with those obtained in density functional theory enables us to conclude that insertion of a limited amount of carbon atoms into the atomic structure stabilises $\alpha$~boron with respect to both carbon-doped $\beta$~boron and boron carbide between 1474~K and 2273~K at 5~GPa, thus extending the stability of (C-doped) $\alpha$~boron beyond the known boundaries of the actual phase diagram. Moreover, the (B$_{11}$C$^{\textit{p}}$) defect predicted by the calculations produces a $n$-type shallow defect level potentially interesting for electronic applications. 

Finally, we show that the higher the formation temperature of $\alpha$~boron at 5~GPa (up to 2273~K), the smaller the carbon content, thus proving that the synthesis temperature is an easy means to  
 control the carbon concentration on demand to produce $n$-type or intrinsic $\alpha$~boron. This finding paves the way, for the first time, to an economically viable mass production of $\alpha$ boron.

\section*{Acknowledgement}

The authors thank Keevin Béneut, Ludovic Delbes, Antoine Jay, Hicham Moutaabbid, Olivier Hardouin Duparc, Guido Roma and Silvia Pandolfi for useful discussions. The X-ray diffraction platform and Raman spectroscopy platform in IMPMC is also acknowledged, as well as SOLEIL synchrotron for the provision of synchrotron radiation facilities. The calculations have been done using \textsc{Quantum ESPRESSO}~\cite{Giannozzi:2009,Giannozzi:2017}. The access to high performance computing resources was granted by the French HPC centers GENCI-IDRIS, GENCI-CINES and GENCI-TGCC (Project 2210) and by the École Polytechnique through the 3L-HPC project. Financial supports from the DIM SIRTEQ (région Île de France), the DGA (France) and the program NEEDS-Matériaux (France) are gratefully acknowledged. The PhD fellowships of A. Chakraborti and Y. Cho have been provided by the Ecole Doctorale of the Institut Polytechnique de Paris, \textit{via}, for YC, the ANR-21-CMAQ-002 project. 




\bibliographystyle{IEEEtran} 

\begin{thebibliography}{10}
\bibitem{Cho:Note:2022:xc_discontinuity}
The discontinuity of exchange and correlation potential is still present in
  {HSE06}.
\bibitem{Cho:Note:2022:slope_lin_reg_Vol}
The slopes of linear regressions were obtained as the following:
  $-$0.011~{[\AA$^3/$at.~\%~C]} for the {(B$_{11}$C$^{\textit{p}}$)} defect;
  0.016~{[\AA$^3/$at.~\%~C]} for the {(B$_{11}$C$^{\textit{e}}$)} defect;
  0.080~{[\AA$^3/$at.~\%~C]} for the {(B$_{12}$)C$^{\textit{Td}}$} defect;
  0.071~{[\AA$^3/$at.~\%~C]} for the
  {(B$_{12}$)C$^{\textit{Td}}\square$C$^{\textit{Td}}$} defect;
  0.024~{[\AA$^3/$at.~\%~C]} for the
  {(B$_{11}$C$^{\textit{p}}$)C$^{\textit{Td}}$} defect;
  0.027~{[\AA$^3/$at.~\%~C]} for the
  {(B$_{11}$C$^{\textit{p}}$)C$^{\textit{Td}}\square$C$^{\textit{Td}}$} defect;
  and 0.036~{[\AA$^3/$at.~\%~C]} for the {(B$_{11}$C$^{\textit{p}}$)CBC}
  defect.
\bibitem{Cho:Note:2022:linear_fit_relative_Vol}
The slope of linear regression for the {(B$_{11}$C$^{\textit{p}}$)} defect was
  that of the note~\cite{Cho:Note:2022:slope_lin_reg_Vol} divided by $V_0$.
\bibitem{Cho:Note:2022:def}
The least energetic defect is the (B$_{12}$)C$^{\textit{Td}}$ defect, in the
  sense that the formation enthalpy has the smallest net positive value.
\end{thebibliography}





\end{document}